\documentclass[journal,fleqn,onecolumn]{IEEEtran}
\usepackage[T1]{fontenc}
\usepackage{cite}
\usepackage{amsmath,makecell,bm}
\usepackage{array,graphics,graphicx,subfigure}
\usepackage{cleveref}
\usepackage{algorithm}  
\usepackage{color,soul}
\newtheorem{theorem}{Theorem}
\usepackage{algorithmic}
\usepackage{multirow,booktabs,setspace}
\usepackage{amsmath,float} 
\allowdisplaybreaks[4]
\setulcolor{black}
\begin{document}
\title{Distributed Urban Freeway Traffic Optimization Considering Congestion Propagation}
\author{Fengkun~Gao, 
        Bo~Yang,~\IEEEmembership{Senior Member,~IEEE,}
        Cailian~Chen,~\IEEEmembership{Member,~IEEE,}\\
        Xinping~Guan,~\IEEEmembership{Fellow,~IEEE,}
        Yang~Zhang
\thanks{This work was supported by NSF of China under Grants 61731012.}
\thanks{F. Gao, B. Yang \textit{(Corresponding author, e-mail: bo.yang@sjtu.edu.cn)}, C. Chen and X. Guan aare with the Department of Automation, Shanghai Jiao Tong University, Shanghai 200240, China, the Key Laboratory of System Control and Information Processing, Ministry of Education of China, Shanghai 200240, China, and also with Shanghai Engineering Research Center of Industrial Intelligent Control and Management, Shanghai 200240, China. }
\thanks{Y. Zhang is with Shanghai Municipal Urban-Rural Construction and Transportation Development Institute.}
}


\maketitle
\begin{spacing}{2.0}
\begin{abstract}
  Traffic optimization strategies are imperative for improving the performance of transportation networks. Most traffic optimization strategies only depend on traffic states of congested road segments, where congestion propagation is neglected. Therefore, we propose a distributed traffic optimization strategy for urban freeways considering the potential congested road segments caused by congestion propagation, called \textit{potential-homogeneous-area} (PHA).
 {Utilizing the historical traffic density data,} we \textcolor{black}{firstly quantify the effect of congestion propagation and} identify \textcolor{black}{PHA} by applying the proposed spatio-temporal lambda-connectedness method. \textcolor{black}{Meanwhile, dynamic capacity constraints of PHA are determined} and are integrated with cell transmission model (CTM) in a centralized traffic optimization problem. To reduce computational complexity and improve scalability, \textcolor{black}{we then propose a double-consensus-based alternating direction method of multipliers algorithm (DC-ADMM) to handle the neighbor coupling constraints and global coupling constraints for solving the problem in a fully distributed way.} We prove that the proposed DC-ADMM algorithm converges to the optimal solution \textcolor{black}{in the condition of convex objective function.} \textcolor{black}{ Finally, simulations based on real traffic density data, collected in Inner Ring Road, Shanghai, China, reveal the effectiveness of our proposed strategy.}

\end{abstract}
\section{Introduction}
\IEEEPARstart{W}ITH cities expanding, urban freeway plays a more and more important part in urban traffic networks. Taking Hangzhou, China, as an example, urban freeway constitutes only about $5\%$ of whole traffic networks but delivers over $30\%$ of all traffic flow \cite{chenadaptive}. As a result, congestion frequently occurs in urban freeways, which may cause economic losses and environmental pollution. To mitigate traffic congestion and improve safety, efficient traffic optimization strategies for urban freeways are imperative.
\begin{table}[t]
  \centering
  \caption{NOMENCLATURE}
  \begin{tabular}{lllll}
  \label{notation}
    \; \textbf{Symbol} &\qquad \qquad\textbf{Name} \\
    \qquad$q_i$ & on-ramp queue length of vehicles\\
    \qquad$\rho_i$ & traffic density of cell $i$ \\
    \qquad$\sigma_i$ & external traffic demand of on-ramp $i$\\
    \qquad$\phi_i$ &traffic flow out of cell $i$ \\
    \qquad$\phi_N$ &traffic flow out of sink cell $N$ \\
    \qquad$r_i$ &on-ramp flow entering cell $i$ \\
    \qquad$L_i$ & length of mainline $i$\\
    \qquad$\omega_i$ &congestion wave speed of cell $i$ \\
    \qquad$v_i$ & free flow speed of cell $i$\\
    \qquad$\beta_i$ &the split ratio of cell $i$ \\
    \color{black}
    \qquad$D_i$ &\color{black} total traffic demand of cell $i$ \\
    \qquad$S_i$ &traffic supply of cell $i$ \\
    \qquad$\rho_i^{\text{max}}$ & maximum traffic density of cell $i$ \\
    \qquad$\phi_i^{\text{max}}$   &maximum traffic flow out of cell $i$ 
  \end{tabular}
  \vspace{-0.6cm}
\end{table}

\subsection{Motivations}
\textcolor{black}{Traffic optimization has attracted much attention and has made considerable progress.} \textcolor{black}{Ramp metering and variable speed limit (VSL) control are the most commonly used manners in freeway traffic optimization,} which have been proved to be effective in avoiding congestion and improving transportation efficiency based on both macro-simulation and micro-simulation \cite{VSLmacrosim, VSLmicrosim}. Ramp metering regulates the number of vehicles entering the mainline passing through on-ramps, 
which is a feedback controller and adjusts metering rate according to real-time traffic states (\textit{e.g.}, flow speed and traffic density of mainlines) \cite{ALINEA1997,ALINEA2003,FeedbackALINEA2013}. 
Variable speed limit control determines the optimal speed for vehicles driving on mainlines to smooth the traffic flow and improve the throughput of freeways. More often, ramp metering and VSL control are combined to achieve better management \cite{VSLRM2005, VSLRM2011, VSLRM2017}.

However, few studies consider the congestion propagation and strong coupling between adjacent road segments when developing traffic control strategies. The fact is that the congested segments have a severe effect on the operation of upstream and downstream, \textit{e.g.}, spillback. These affected segments, \textcolor{black}{\textit{i.e., potential-homogeneous-area} (PHA),} have a possibility of getting congested due to the congestion propagation. 

\textcolor{black}{The above observations motivate us to design a distributed traffic optimization strategy considering the effect of congestion propagation. The challenges mainly lie in two aspects. The first is how to analyze the mechanism of congestion propagation and identify PHA accurately. Because it is hard to quantify the magnitude and the range of congestion propagation. When considering traffic congestion propagation, both neighbor coupling constraints and global coupling constraints are incorporated into the traffic optimization problem. Therefore, the second is how to exploit efficient distributed optimization algorithms to deal with the coupling constraints such that the scalability and resilience of intelligent transportation systems can be guaranteed.}

\color{black}
\subsection{Background and Related Works}
\color{black}
\label{relatedwork}
\textcolor{black}{The proposed strategy mainly embraces three aspects, coordination of ramp metering and VSL control, congestion propagation identifying and distributed traffic optimization.}

\textcolor{black}{\textbf{Coordination of Ramp Metering and VSL Control:} Recently, based on the coordination of ramp metering and VSL control, both centralized and distributed traffic optimization strategies have been developed.} Centralized strategies, such as feedback-based \cite{centralized2} and predictive-based \cite{Centralized1,Centralized3}, have access to the global optimum and are convenient to implement. Nonetheless, the above centralized strategies have several detrimental issues. Due to the requirement of global data collection, centralized control is not scalable and suffers substantial computation and communication burdens. Besides, single control and data center (\textit{e.g.}, cloud) is not resilient for control failures. These problems can be avoided by \textcolor{black}{introducing} distributed traffic control methods, in which a sizeable centralized problem is divided into several subproblems. Each small subproblem can be solved locally by an agent (\textit{e.g.}, roadside unit) relying on communication with neighbors, \textcolor{black}{as shown in the studies \cite{PAN2021102987,DMPC2,9042890,wu2020differential}.} \textcolor{black}{Additionally, with the rapid increase of electric vehicles, ITS has the potential of decreasing the operation cost of smart grids by scheduling the EVs with charging demand \cite{8928490,8886346,8669821}.}

\textcolor{black}{
  \textbf{Congestion Propagation Identifying:} 
  Due to the mobility of vehicles, congestion propagation occurs frequently and has a severe effect on the performance of transportation networks. In \cite{NRR2016,CongestionPropagation2}, the cost of both congestion and congestion propagation is taken into account, and an effective routing algorithm is designed. In \cite{CongestionPropagation1}, a histogram-based model for congestion detection and congestion propagation is proposed and rerouting strategy is also investigated, which takes precaution measures before the critical point of congestion occurrence.
}

\textcolor{black}{
  \textbf{Distributed Traffic Optimization:}
  The methods can be roughly divided into two types, model-based and model-free. Literally, model-based methods require accurate and effective traffic models to formulate the traffic dynamics, \textit{e.g.}, cell transmission model (CTM, presented in section \ref{sectionCTM}) and METANET \cite{METANET}. In \textcolor{black}{\cite{DisSystemofSystem,DistributedRM,ShareStates},} the whole freeway system is divided into subnetworks and several distributed optimization methods are designed, such as distributed model predictive control (DMPC) and asynchronous edge-based alternating directions method of multiplier (ADMM) algorithm. Model-free methods, such as Q-learning \cite{DRL}, rely only on a simulation environment to develop strategies, while no consideration of prior knowledge leads to poor performance during initial learning process.}

  \textcolor{black}{The current works have extensively developed distributed optimization methods and have taken congestion propagation into account, while the lack of quantification methods for congestion propagation hinders the performance improvement. Besides, the optimization methods do not integrate with the analysis of congestion propagation.}
\subsection{Contributions}
\textcolor{black}{In this paper, we aim to propose a method to quantify the effect of congestion propagation utilizing the information of road network topology and historical traffic data. The quantification results are incorporated into a traffic optimization problem. Additionally, a distributed algorithm is designed to solve the centralized problem for reducing computational complexity and improving scalability, where coupling constraints are handled.} The main contributions are summarized as follows:
\begin{enumerate}
  \item \textcolor{black}{A spatio-temporal lambda-connectedness method is proposed to \textcolor{black}{quantify the effect of congestion propagation and} identify PHA}, which is based on analyzing historical traffic data. \textcolor{black}{Meanwhile, dynamic capacity constraints of PHA are determined and are integrated into the finite-time horizon optimization problem.}
  \item \textcolor{black}{A double-consensus-based alternating direction method of multipliers algorithm (DC-ADMM)} is designed to solve the \textcolor{black}{traffic optimization problem in a fully distributed way}. We also prove that the proposed algorithm converges to the globally optimal solution in the condition of convex objective function.
  \item Simulations are designed to evaluate the performance of the proposed strategy on the real traffic data \textcolor{black}{collected in Inner Ring Road,} Shanghai, China. The results demonstrate that our proposed strategy can reduce congestion and improve traffic efficiency significantly.
\end{enumerate}

\subsection{Organization}
The remainder of the paper is organized as follows. Section~\ref{Problem Statement} demonstrates the \textcolor{black}{overview } \textcolor{black}{of the proposed method and the formulations of traffic dynamics.} \textcolor{black}{The PHA identification and the formulation of finite time horizon traffic optimization problem} are introduced in Section~\ref{Centralized}. Section~\ref{Distributed} gives the details of \textcolor{black}{the proposed DC-ADMM} algorithm to solve the problem. Section~\ref{simulation} shows simulation results based on real traffic density data \textcolor{black}{collected in Inner Ring Road,} Shanghai, China. Section~\ref{conclusion} summarizes the work and discusses the future work we will concentrate on. 

\section{System description and modeling}
\label{Problem Statement}
In this section, we \textcolor{black}{provide the overview of the proposed traffic optimization strategy,} and introduce the formulations of traffic dynamics, \textit{i.e.}, cell transmission model (CTM). 

\begin{figure}[t]
  \vspace{-0.3cm}
  \centering
  \includegraphics[scale=0.3]{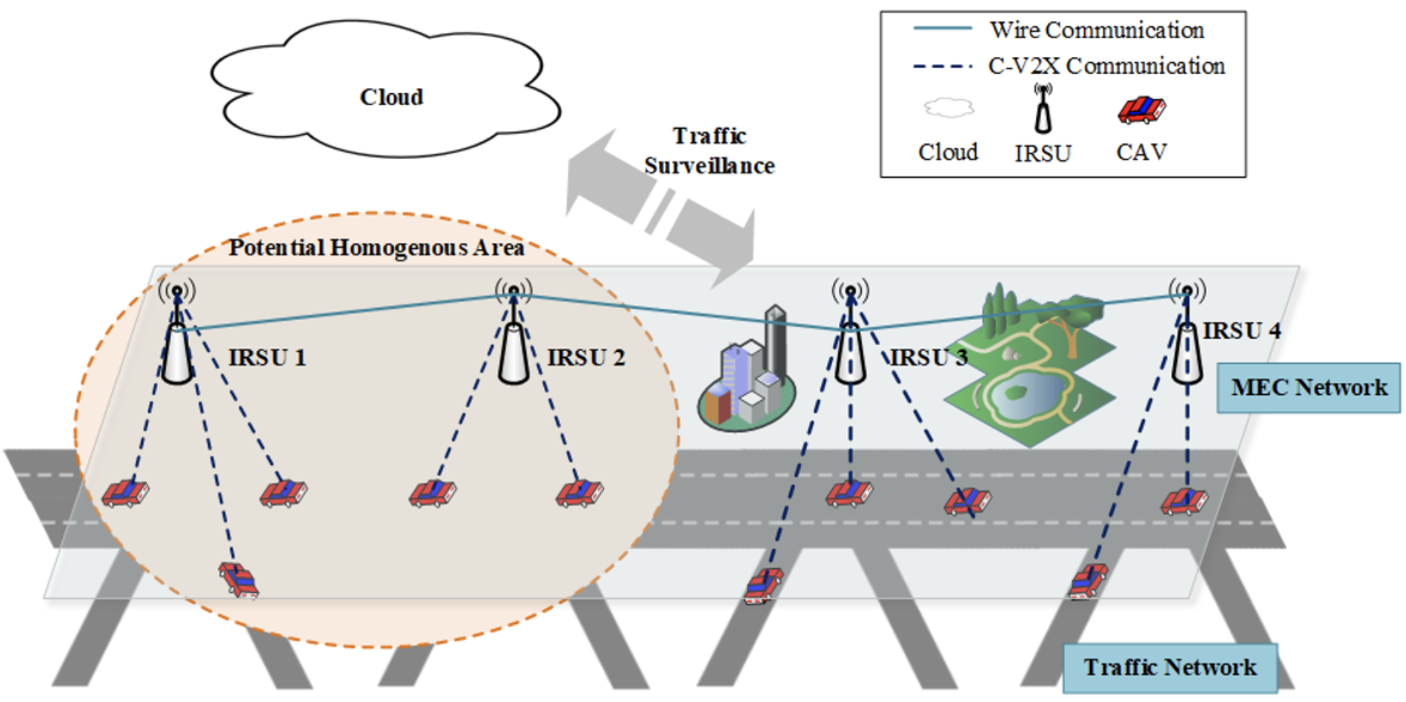}
  \vspace{-0.3cm}
  \caption{Intelligent transportation system enhanced by MEC and C-V2X.}
  \label{ITS}
\end{figure}

\color{black}
\subsection{Overview}
\subsubsection{Scenario and Assumptions}

\textcolor{black}{Intelligent transportation system is considered a promising way to mitigate traffic congestion, improve transportation efficiency and enhance travel safety. However, ITS generates massive amounts of traffic data every time, whose collecting and processing require substantial communication, storage, and computing resources.}
\color{black}

\textcolor{black}{In our work, we consider the following scenario of ITS enhanced by MEC and C-V2X. As shown in Fig. \ref{ITS}, the system consists of connected and automated vehicles (CAVs), intelligent road-side units (IRSUs), and the cloud, \textit{i.e.}, traffic management center.} C-V2X technology satisfies the requirements of both safety and entertainment applications for low-latency, high-reliability, and high-throughput communications \textcolor{black}{\cite{CV2X,7572192}} \textcolor{black}{, which has been tested in application level\cite{8962283}.} MEC achieves powerful computing capabilities at the edge of the network toward computational-intensive tasks \textcolor{black}{\cite{MEC,8933514} and edge intelligence \cite{9123504}}. CAVs are assumed to be equipped with C-V2X communication devices and can upload their states, \textit{e.g.}, velocity, location, and origin-destination (OD) information, to IRSUs and cloud periodically. IRSUs are integrated with base stations, MECs, and sensors (\textit{e.g.}, cameras), which can provide perception, communication, computation, and storage services. 

We assume that the whole freeway network can be divided into several subnetworks, and each subnetwork is deployed with an IRSU. Hence, the global traffic state can be detected by \textcolor{black}{aggregating the data collected at the edge nodes.} Those time-sensitive and computational-intensive tasks can \textcolor{black}{also} be processed by MEC to avoid high communication and computation delay of cloud computing. Moreover, each IRSU can exchange information with neighbor IRSUs to achieve coordinated operation, \textit{e.g.}, the implementation of distributed traffic optimization algorithms.

\subsubsection{\textcolor{black}{The Architecture of The Proposed Strategy.}}
\begin{figure}[t]
  \vspace{-0.3cm}
  \centering
  \includegraphics[scale=0.35]{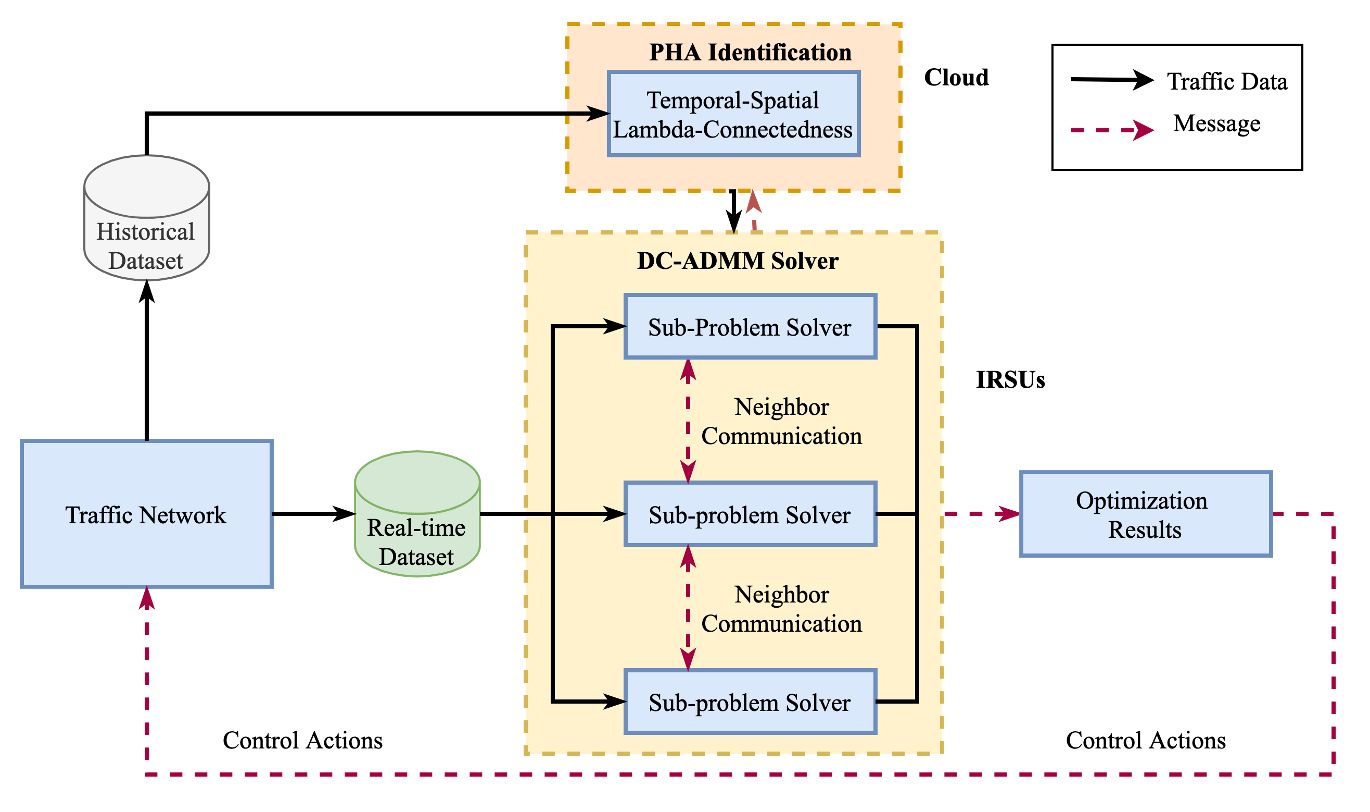}
  \vspace{-0.4cm}
  \caption{The architecture of the proposed traffic optimization strategy}
  \label{scheme_architecture}
\end{figure}
\textcolor{black}{
The architecture and workflow of the proposed traffic optimization strategy are illustrated in Fig. \ref{scheme_architecture}. In aforementioned discussions, the collected historical traffic data are stored in the cloud for traffic surveillance. When congestion occurs, these historical data corresponding to the congested segments is utilized to identify PHA based on our proposed spatio-temporal lambda-connectedness method in the cloud. The identification results are then sent to those IRSUs deployed in PHA, \textcolor{black}{as shown in Fig. \ref{ITS}}. At the same time, DC-ADMM algorithm is started with cooperations of IRSUs to obtain the optimal traffic control strategies, which are sent to CAVs to achieve the coordination of ramp metering and variable speed limit control until congestion disappears.
}
\color{black}
\subsection{Cell Transmission Model}
\label{sectionCTM}
\begin{figure}[htbp]
  \vspace{-0.5cm}
  \centering
  \includegraphics[scale=0.35]{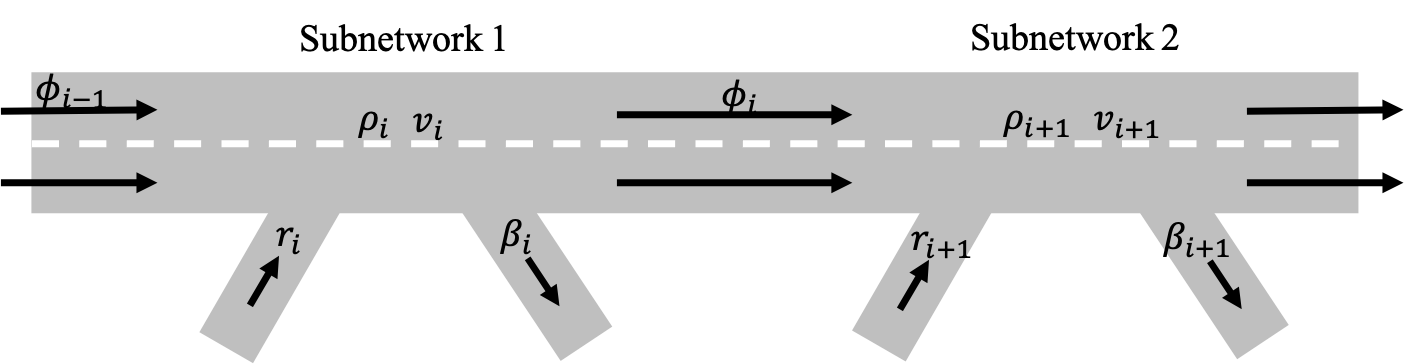}
  \vspace{-0.3cm}
  \caption{Demonstration of freeway structure.}
  \label{freeway}
  \vspace{-0.3cm}
\end{figure}
To interpret traffic dynamics, we introduce a commonly used model, widely known as the cell transmission model (CTM), which is the extension of flow conservation law. 

As shown in Fig. \ref{freeway}, the urban freeway network consists of mainlines and ramps (on-ramps and off-ramps). Each road segment can be seen as a cell that vehicles drive in and out. 
The CTM is formulated as follows: 
\begin{equation}
  q_i(k+1)=q_i(k)+\Delta t \big(\sigma_i(k)-r_i(k)\big) 
  \label{queue of ramp}
\end{equation}
\begin{equation}
  \label{mainline density}
  \begin{split}
    \rho_i(k+1)=\rho_i(k)+\frac{\Delta t}{L_i}\big(  &\phi_{i-1}(k)(1-\beta_{i-1}(k))\\
    &+r_{i}(k)-\phi_i(k)\big)
  \end{split}
\end{equation}
\begin{equation}
  \phi_i(k)\!=\!\min\{v_i(k)\rho_i(k),\! \phi_i^{\text{max}}\}\frac{\min\{D_i(k),S_{i+1}(k)\}}{D_i(k)}
  \label{cell flow}
\end{equation}
\begin{equation}
  \phi_N(k)=\min\{v_N(k)\rho_N(k), \phi_N^{\text{max}}\} 
\end{equation}
\begin{equation}
  r_i(k)=d_i\frac{\min\{D_i(k),S_{i+1}(k)\}}{D_i(k)} 
\end{equation}
\begin{equation}
  d_i(k)=\min\{q_i(k),c_i(k)\} 
\end{equation}
\begin{equation}
  D_i(k)=\min\{v_i(k)\rho_i(k), \phi_i^{\text{max}}\}\big(1-\beta_i(k)\big)+d_i(k) 
\end{equation}
\begin{equation}
  S_i(k)=\min\{\omega(\rho_i^{\text{max}}-\rho_i(k)), \phi_i^{\text{max}}\},
  \label{cell capacity}
\end{equation}
where the \textcolor{black}{traffic demand $D_i$ represents those vehicles that have the requirement of passing through road segment $i$ to arrive at the destination.} The corresponding notations are presented in Table \ref{notation} and the sample interval is assumed to be $\Delta t$. The evolution of traffic dynamics can be updated by calculating the above formulations (\ref{queue of ramp})-(\ref{cell capacity}). However, constraints (\ref{cell flow})-(\ref{cell capacity}) are non-convex, which leads that the traffic optimization problem is hard to achieve an efficient solution. To tackle the problem, non-convex constraints can be relaxed as:
  \begin{equation}
    \phi_{i}(k) \leq \min \left\{\rho_{i}(k) V^{\max }, \phi_{i}^{\max }\right\} \label{e9}
  \end{equation}
  \begin{equation}
    \label{e10}
    \begin{split}
      &\phi_i(k)\left(1-\beta_{i}(k)\right)+r_{i}(k)  \\ 
      &\qquad \leq \min \left\{\phi_{i+1}^{\max }, \omega\left(\rho_{i+1}^{\max }-\rho_{i+1}(k)\right)\right\} 
    \end{split}
  \end{equation}
  \begin{equation}
    r_{i}(k) \leq \min \left\{q_{i}(k), C^{\max }\right\} \label{e11},
  \end{equation}
where $V^{\text{max}}$ and $C^{\text{max}}$ are the upper bound of $v$ and $c$, respectively. The results in \cite{ShareStates} and \cite{OptimalACC} show that optimal solution of original problem can be obtained by solving the relaxed optimization problem with inequality constraints (\ref{e9})-(\ref{e11}) such that the \textcolor{black}{optimal} ramp metering rate $r$ and flow speed $v$ can be constructed. 

Further, we let $x_{i}(k)=\left[\rho_{i}(k) \quad q_{i}(k)\right]^{\top}$ represent the local traffic states of subnetwork $i$ at discrete time step $k$ and $u_{i}(k)=\left[r_{i}(k) \quad \phi_{i}(k)\right]^{\top}$. The uncontrollable traffic flow (\textit{i.e.}, traffic demand from external networks) is defined as $\psi_i(k)=\left[0 \quad \sigma_i(k) \right]$. We suppose the entire network can be divided into $M$ subnetworks, which is denoted by the set $\mathcal{M}$. The connectivity of each network can be described by a graph $(\mathcal{M},\mathcal{E})$, where $\mathcal{E}$ represents the traffic flow direction. Traffic flow from subnetwork $i$ towards subnetwork $j$ is represented by $(i,j) \in \mathcal{E}$. For each subnetwork, the traffic dynamics (\ref{queue of ramp}) and (\ref{mainline density}) can be reformulated as follows:
  \begin{equation}
    x_i(k+1)=x_i(k)+B_{ii}u_i(k)+B_{ij}u_j(k)+\psi_i(k),
  \label{cell state}
  \end{equation}
where $B_{ii}$ and $B_{ij}$ are suitable defined matrices and $u_j(k)$ accounts for the influence of adjacent subnetworks. 

\section{Centralized formulation}
\label{Centralized}
In this section, we firstly introduce the proposed spatio-temporal lambda-connectedness method to identify \textcolor{black}{PHA}. \textcolor{black}{Meanwhile,} dynamic capacity constraints of \textcolor{black}{PHA} is \textcolor{black}{also determined.} Finally, a centralized \textcolor{black}{finite-time horizon} traffic optimization problem is formulated.

\subsection{Spatio-Temporal Lambda-Connectedness Method}
When traffic congestion occurs, restricting traffic flow entering the congested area is a commonly used manner, which is achieved by ramp metering in freeway networks. Most of the existing work only considers the dynamic capacity limitation and traffic restriction of congested segments. However, the fact is that traffic congestion will spread upstream and downstream due to the strong coupling between adjacent segments and the mobility of traffic flow \cite{LZHVTC}. For this reason, potential congested segments, \textit{i.e.}, \textcolor{black}{PHA}, should be taken into account when developing traffic optimization strategies. Due to the complexity of network topology and the spatio-temporal variability of traffic flow, it is challenging to identify \textcolor{black}{PHA}. To tackle this problem, we propose a spatio-temporal lambda-connectedness method, which is the expansion of lambda-connectedness. This method quantifies the temporal and spatial relationships between those segments in \textcolor{black}{PHA}.

\begin{figure}[htbp]
  \centering
  \subfigure[Road Network]{
    \label{roadsegment}
    \begin{minipage}[t]{0.35\linewidth}
    \includegraphics[width=1\linewidth]{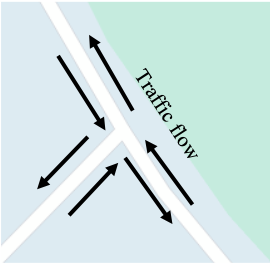}
  \end{minipage}%
  } \quad
  \subfigure[Linkage Network]{
    \label{linkage}
    \begin{minipage}[t]{0.45\linewidth}
    \includegraphics[width=1\linewidth]{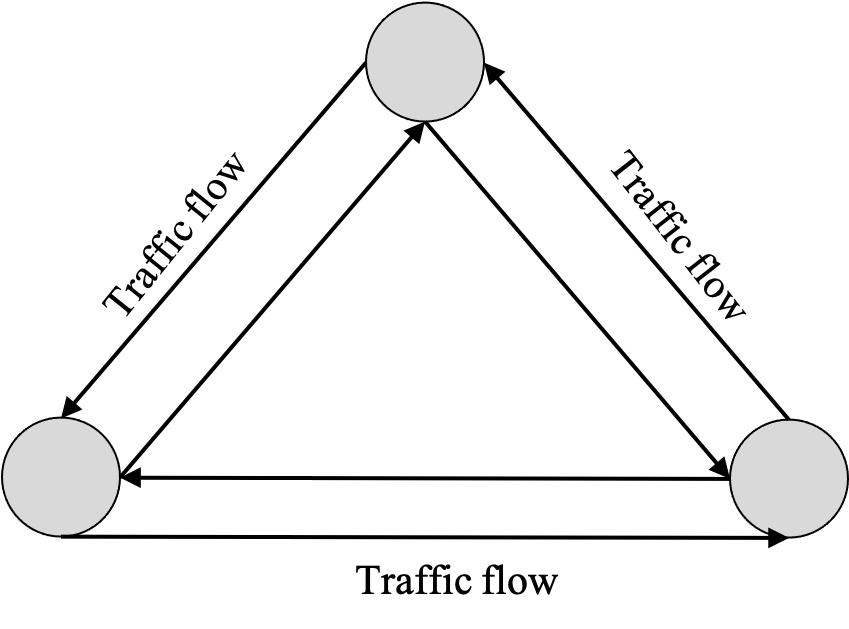}
  \end{minipage}
  }
  \vspace{-0.3cm}
    \begin{center}
      \caption{Road network and linkage network.}
      \label{linkagenet}
    \end{center}
    \vspace{-0.6cm}
\end{figure}

Lambda-connectedness is developed from graph theory, which can describe partial connectivity and fuzzy relation of two vertices in a discrete space. In order to apply this method to traffic systems, we use a linkage network to describe the topology of traffic networks, such as Fig. \ref{linkagenet}. The road network Fig. \ref{roadsegment} can be translated into a linkage network Fig. \ref{linkage}. In a linkage network, each vertex represents a cell (segment), and edges describe the mobility paths of traffic flow between neighbor cells. 

We start with defining a potential function $\alpha_{\rho,t}\left(\overline{x},\overline{y}\right)$ to measure the neighbor-connectivity of $\overline{x}$ and $\overline{y}$ at time slot $t$. 
\begin{equation*}
  \alpha_{\rho,t}\left(\overline{x},\overline{y}\right)=a / e^{\left|\rho_{\overline{x}}\left(t\right) \ - \ \rho_{\overline{y}}\left(t\right)\right|} + b / e^{|d(\overline{x},\overline{y})|},
  \label{e13}
\end{equation*}
where $\rho_{\overline{x}}\left(t\right)$ and $\rho_{\overline{y}}\left(t\right)$ are traffic density of segment $\overline{x}$ and $\overline{y}$ at time slot $t$, respectively. $d(\overline{x},\overline{y})$ is the geographic distance between $\overline{x}$ and $\overline{y}$. $a$ and $b$ are adjustable weighted parameters. Because of the fact that those segments in \textcolor{black}{PHA} may have similar traffic dynamics (density, velocity, etc.), we choose traffic density and geographical distance to construct the potential function. The similar manipulation can be found in \cite{NetPartition}, where traffic density and lambda-connectedness method are used to partition network into subnetworks with distinct macroscopic fundamental diagram (MFD) properties.

In graph theory, a path $\pi\left(\overline{x}_1,\overline{x}_n\right)$ is defined by a finite sequence $\{\overline{x}_1,\overline{x}_2,\ldots,\overline{x}_n\}$. In our work, we consider the temporal variability of traffic state. For two vertices $\overline{x}_1$ and $\overline{x}_n$, a path at sample time $t$ is represented by $\pi_t\left(\overline{x}_1,\overline{x}_n\right)$, which is obtained by analyzing the traffic data of time slot $t$. The path-connectivity $\beta_{\rho,t}$ of a path $\pi_t\left(\overline{x}_1,\overline{x}_n\right)$ is defined as \textcolor{black}{the following equation:}
  \begin{equation*}
    \beta_{\rho,t}\left(\pi_t\left(\overline{x}_{1}, \overline{x}_{n}\right)\right)=\min \left\{\alpha_{\rho,t}\left(\overline{x}_{i}, \overline{x}_{i+1}\right) \mid i=1, \ldots, n-1\right\}.
    \label{e14}
    \end{equation*}

Further, analyzing all path-connectivity $\beta_{\rho,t}\left(\pi_t\left(\overline{x}_{1}, \overline{x}_{n}\right)\right)$, the degree of connectedness of two vertices is defined as:
  \begin{equation*}
    \begin{split}
      C_{\rho}\left(\overline{x}_1, \overline{x}_n\right)=\max \big\{ \beta_{\rho,t} &\left(\pi_t\left(\overline{x}_1, \overline{x}_n\right)\right) \mid \text { for all path } \pi_t \big \}.
    \end{split}
    \label{e15}
  \end{equation*}

For a given $\Lambda \in [0,1]$, $\overline{x}_{1}$ and $\overline{x}_{n}$ are said to be lambda-connectedness if the following inequality holds
\begin{equation*}
  C_{\rho}\left(\overline{x}_1, \overline{x}_n\right) \geq \Lambda.
  \label{e17}
\end{equation*}

Following the above steps, the \textcolor{black}{range of PHA} can be identified. All cells in this area are lambda-connectedness with the congested ones. The historical data at different moments are sampled for identifying the \textcolor{black}{PHA}, which implies that both spatial and temporal information of traffic dynamics are considered in the formulation.
\vspace{-0.3cm}
\subsection{Capacity Constraint of \textit{Potential-Homogeneous-Area}}
Based on the analysis above, we can conclude that segments in \textcolor{black}{PHA} have similar traffic dynamics and work like a single cell with multi-input and multi-output. In this regard, it is reasonable to take this area as a whole when developing traffic control strategies. Similar to a single cell, the transportation capacity of this area is upper bounded and fluctuates with time, which can be described by the dynamic capacity constraint
\begin{equation}
  \label{globalconst}
  \sum_{i=1}^{M} \bar{f}_i\left(u_i(k)\right) \leq \bar{\delta}_t \quad \forall k \in T,
\end{equation}
where $\bar{\delta}_t$ is the transportation capacity (in vehicles per hour, or veh/h) of \textcolor{black}{PHA} consisting of $M$ subnetworks, which varies with time $t$. The value of $\bar{\delta}_t$ can be affected by road conditions and external environments, \textit{e.g.}, weather conditions and accidents. At the moment of $t$, it is specified by the cloud based on analyzing both the historical and real-time data. Moreover, it is assumed that $\bar{\delta}_t$ remains the same in a short time horizon $T$. The assumption is rational because the traffic capacity is not time-sensitive. $\bar{f}_i(u_i(k))$ is defined as a function of external traffic demand. Specifically, it can be the linear combination of external traffic demand, \textit{e.g.}, $\bar{f}_i(u_i(k))=h_i u_i(k)$. $h_i$ is the corresponding weighted parameter, which can be adjusted according to the priority of road segments. 

\vspace{-0.2cm}
\subsection{Centralized Traffic Optimization Problem}
To optimize the urban freeway, a proper objective function is essential. A widely used performance index is total travel time (TTT), which is defined as:
\begin{equation}
  J_{T}=\sum_{i=1}^{M} \sum_{k=0}^{N-1}(\Delta t)\big(q_{i}(k)+\rho_{i}(k)L_i\big),
  \label{objective}
\end{equation}
where $q_i$ is the vehicle queue length of on-ramp $i$, $\rho_i$ is the traffic density of mainline $i$, and $k$ is the discrete time-step. There are also other indexes used for traffic control, such as total travel distance (TTD), total delay (TD), and density balancing \cite{OptimalACC,DisLargeScale}.

Considering the CTM and the dynamic capacity constraint, the \textcolor{black}{finite time horizon} traffic optimization problem is formulated in a centralized form as follows:
\begin{equation*}
  \begin{split}
    \mathrm{P1} : & \min _{u_i} \sum_{i=1}^{M} \sum _{k=0}^{N-1} (\Delta t)\big(q_{i}(k)+\rho_{i}(k)L_i\big) \\
    \text { s.t. } & x_i(k+1)=x_i(k)+B_{ii} u_ i (k)+B_{ji} u_j(k)+\psi_i(k)  \\
    & \sum_{i=1}^{M} \bar{f}_i\left(u_i(k)\right) \leq \bar{\delta}(t),\quad (\ref{e9})-(\ref{e11})\\
    & \text{for} \quad i=1, \cdots, M \ ; \  k=0, \cdots, N-1.
  \end{split}
\end{equation*}

When congestion occurs at the moment of $t$, \textcolor{black}{the PHA is first identified based on spatio-temporal lambda-connectedness method, and} dynamic capacity constraint (\ref{globalconst}) is determined. The above finite time horizon optimization problem P1 is then solved to develop the traffic optimization strategies. These instructions of ramp metering and the speed limit are finally sent to CAVs (V2I) and traffic signs to achieve coordinated traffic control. 

However, the centralized problem is difficult to solve efficiently with the increase of network size. Besides, it suffers from high communication and storage costs, because the traffic data need to be uploaded to the cloud, and then the computing results need to be downloaded to the IRSUs or CAVs. In Section~\ref{Distributed}, we proposed a fully distributed algorithm to solve the above centralized optimization problem based on partial augmented Lagrangian and dual-consensus ADMM.

\section{Distributed Solution}
\label{Distributed}
\begin{figure}[t]
  \vspace{-0.3cm}
  \centering
  \includegraphics[scale=0.45]{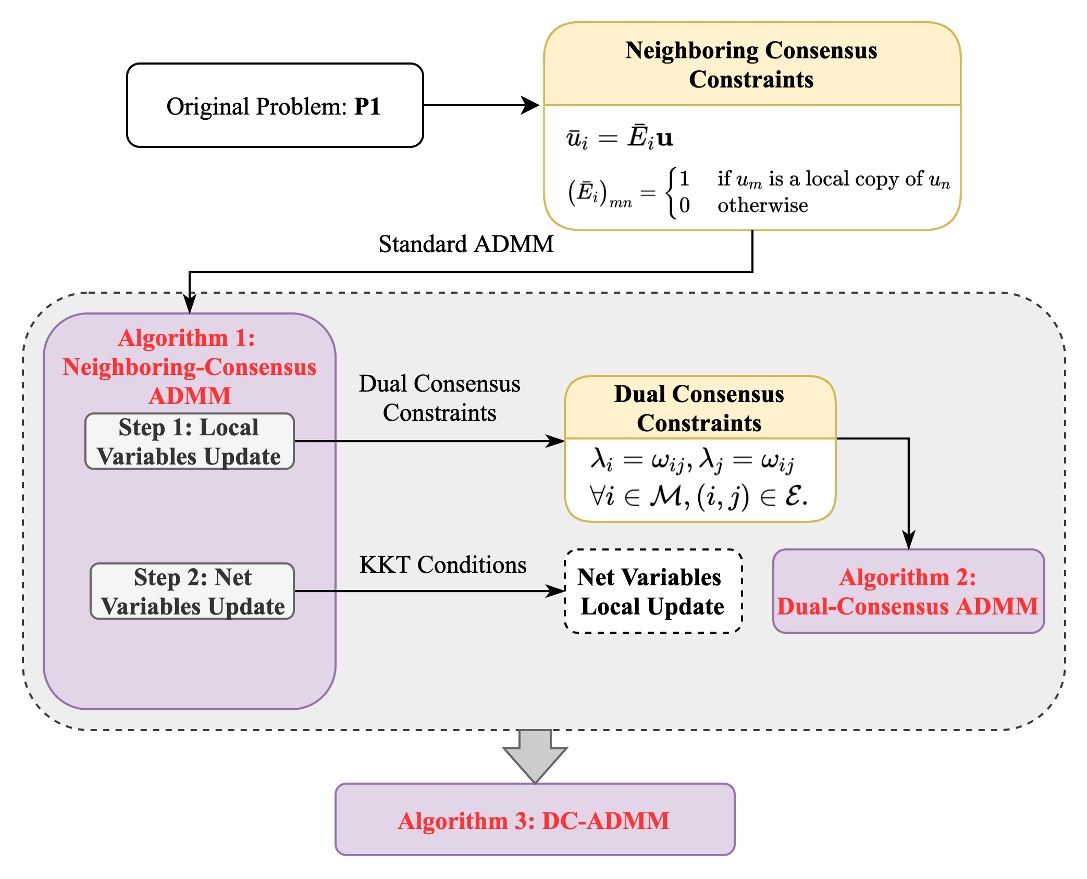}
  \vspace{-0.4cm}
  \caption{The derivation of DC-ADMM.}
  \label{algorithmfig}
\end{figure}

In this section, we clarify the proposed \textcolor{black}{DC-ADMM} algorithm for solving the aforementioned centralized problem. \textcolor{black}{The overall derivations of DC-ADMM are shown as Fig. \ref{algorithmfig}.} Algorithm \ref{algorithm1} is firstly given \textcolor{black}{by introducing neighboring consensus constraints and applying} standard ADMM. To handle the global coupling constraints in updating local variables of Algorithm \ref{algorithm1}, the dual problem is investigated and Algorithm \ref{algorithm2} is obtained. \textcolor{black}{Besides, the derivation based on Karush-Kuhn-Tucker conditions (KKT) reveals that net variables can also be updated locally.} Finally, a fully distributed algorithm is given by Algorithm \ref{algorithm3} base on Algorithm \ref{algorithm1} and Algorithm \ref{algorithm2}. The convergence of Algorithm \ref{algorithm3} is also analyzed theoretically.

\subsection{Neighbor Shared Variables Decouple}
Note that, (\ref{cell state}) indicates that the evolution of local traffic states is influenced by neighbors, and (\ref{globalconst}) is a global constraint. The two issues lead that the distributed algorithm can not be applied directly to P1. In our considering scenario, neighbor subnetworks can exchange information with each other by IRSUs equipped with communication devices. By this means, we assume that each agent (IRSU) can maintain local copies of the \textcolor{black}{neighbor} shared variables, and then enforce them to have the same value by introducing neighboring consensus constraints \cite{DistributedOPF}. 

We denote by ${N}_i$ the set of neighbors of agent $i$, and local copies of shared variables are denoted by $\bar{u}_i$. Hence, the neighboring consensus constraints are given by
\begin{equation}
  \bar{u}_i=\bar{E}_i \mathbf{u},
  \label{consistency}
\end{equation}
where $\mathbf{u}$ denotes the vectorized $(u_i)_{i\in \mathcal{M}}$ called net variables, and $\bar{E}_i$ is given by 
\begin{equation}
  (\bar{E}_{i})_{mn}=\left\{\begin{array}{ll}
    1 & \text { if } u_m \text{ is a local copy of } u_n \\
    0 & \text { otherwise }.
  \end{array}\right.
\end{equation}
Assuming $\hat{u}_i =[u_i \; \bar{u}_i]$ is the concatenations of $u_i$ and $\bar{u}_i$ for $k=0,\dots,N-1$, $\mathrm{P}1$ can be reformulated as 
\begin{equation*}
  \begin{split}
    &\mathrm{P 2} :  \min _{\hat{u}_i,x_i} \sum_{i=1}^{M} J_{i}\left(x_{i}, \hat{u}_{i}\right) \\
    & \text { s.t. }\sum_{i=1}^M f_i(\hat{u}_i) \leq \delta_t, \quad \hat{u}_i= E_i \mathbf{u},\quad \hat{u}_i \in \mathcal {U}_i\\
    & \qquad \text{for}\quad i=1, \cdots, M,
  \end{split}
\end{equation*}
where $f_i(\hat{u}_i)=H_i \hat{u}_i$ and $ E_i$ is the corresponding matrix. The local optimization objective is $J_i(x_i,\hat{u}_i)=\sum _{k=0}^{N-1}(\Delta t)\big(q_{i}(k)+\rho_{i}(k)L_i\big)$. The set $\mathcal{U}_i$ and matrices $H_i$ is defined by 
\begin{equation}
  \label{setU}
  \begin{aligned}
    \mathcal{U}_i:=\Big\{ &x_i(k+1)=x_i(k)+B_i\hat{u}_i+\psi_i(k), \\
    &(\ref{e9})-(\ref{e11}),\qquad \text{for} \quad k=0, \cdots, N-1\Big\}.
  \end{aligned}
\end{equation}
\begin{equation}
  \delta_t=\left[\begin{array}{cccc}
  \bar{\delta}_t & \bar{\delta}_t & \cdots & \bar{\delta}_t 
\end{array}\right]^{\top}
\quad H_i=\left[\begin{array}{c}
  \bar{H}_i\\
  \boldsymbol{0}
  \end{array}
  \right]
\end{equation}
\begin{equation}
  \bar{H}_i=\left[\begin{array}{ccccc}
  h_i & 0 & 0 & \cdots & 0 \\
  0 & h_i & 0 & \cdots & 0 \\
  \vdots & \vdots & \ddots & \ddots & \vdots \\
  0 & 0 & \cdots & 0 & h_i
  \end{array}\right]
\end{equation}

By introducing neighboring consensus constraints, shared variables between neighbor subnetworks are decoupled. However, the global constraints \textcolor{black}{(\ref{globalconst})} are still not eliminated. To handle the problem, we convert $\mathrm{P2}$ into a problem with penalty based on partial augmented Lagrangian of $\mathrm{P2}$.
\begin{algorithm}[tbp]
	\caption{\textbf{Neighboring-Consensus ADMM}}
	\label{algorithm1}
	\begin{algorithmic}
    \STATE \textbf{Initialization:} Set $n=0$, and give $x_i^{0}$, $\hat{u}_i^{0}$, $\theta_i^0$, $\mathbf{u}^0$
    \STATE \textbf{Repeat}
    \STATE \textbf{1) Local variables update:} Update the local variables $(\hat{u}_i^{n+1},x_i^{n+1})$ by solving the following problem:
    \begin{equation*}
      \begin{split}
        \mathrm{P 3} :  \min \  & \sum _{i=1}^M J_i (x_i,\hat{u}_i) + \sum_{i=1}^m \bigg ((\theta_i^{n})^\top (\hat{u}_i-{E}_i \mathbf{u}^{n})\\
      & \qquad \qquad \qquad \qquad \quad + \frac{\rho_1}{2} \left\| \hat{u}_i - {E}_i \mathbf{u}^{n}\right\|_2^2 \bigg )\\
      &\text{s.t.} \quad 
      \sum_{i=1}^M f_i(\hat{u}_i) \leq \delta_t, \quad \; \hat{u}_i \in \mathcal {U}_i.\\
      \end{split}
    \end{equation*}
    \STATE \textbf{2) Net variables update:} The net variable $\mathbf{u}^{n+1}$ can be obtained by solving the following problem P4:
    \begin{equation*}
      \mathrm{P4 :} \min \sum_{i=1}^{M} \left ( (\theta_i^{n})^{\top}(\hat{u}_i^{n+1}-{E}_i \mathbf{u})+ \frac{\rho_1}{2} \left\| \hat{u}_i^{n+1} - {E}_i \mathbf{u}\right\|_2^2 \right ).
    \end{equation*}
    \STATE \textbf{3) Dual variables update:} Each agent updates dual variable $\mathbf{\theta}_i^{n+1}$ as
    \begin{equation*}
       \theta_i^{n+1}=\theta_i^{n}+ \rho _1 \left(\hat{u}_i^{n+1} - {E}_i \mathbf{u} ^{n+1}\right).
    \end{equation*}
    \STATE \textbf{4) Set:} $n=n+1$
    \STATE \textbf{Until} the predefined stopping criterion is satisfied and return $\left(\hat{u}_i^{n+1},x_i^{n+1}\right)$, $\mathbf{u}^{n+1}$, $\theta_i^{n+1}$. 
	\end{algorithmic}
\end{algorithm}

\vspace{-0.3cm}
\begin{equation*}
  \label{PartialLagrangian}
  \mathcal{L}_p = \sum_{i=1}^{M} \left ( J_i (x_i, \hat{u}_i) + \theta_i^{\top}(\hat{u}_i-E_i \mathbf{u})+ \frac{\rho_1}{2} \left\| \hat{u}_i - E_i \mathbf{u}\right\|_2^2 \right ),
\end{equation*}
where $\theta_i$ is the dual variable associated with (\ref{consistency}) and $\rho_1$ is a penalty parameter. Based on the standard ADMM, we have Algorithm \ref{algorithm1}.
The stopping criterion can be defined as satisfying the following two conditions $|\mathcal{R}_{i}^{n+1}|=|\hat{u}_{i}^{n+1}-E_{i}\mathbf{u}^{n+1}| \leq \epsilon_1$ and $|\hat{u}_i^{n+1}-\hat{u}_i^n|\leq\epsilon_2$, where $n$ is the iteration step. Moreover, the maximum iterations can be utilized for the trade-off between accuracy and computation cost.   

However, the steps 2) and 3) of Algorithm \ref{algorithm1} show that the update of local variables and net variables still requires global information, which is not a fully distributed algorithm. In section \ref{Localupdate}, we further propose a distributed algorithm based on dual-consensus ADMM to update local variables just depending on neighbors' communication. Similarly, the derivation in \ref{Netupdate} shows that net variables can also be updated locally just with neighbors' information.

\subsection{Local Variables Update}
\label{Localupdate}
In this section, we will give the method to solve the problem P3 in step 2) of algorithm \ref{algorithm1}. We define 
\begin{equation}
  \begin{split}
    F_i(x_i,\hat{u}_i)=J_i (x_i,\hat{u}_i) &+ (\theta_i^{n})^{\top}(\hat{u}_i-E_i \mathbf{u}^{(n)})\\
    &+ \frac{\rho_1}{2} \left\| \hat{u}_i - E_i \mathbf{u}^{(n)}\right\|_2^2.
  \end{split}
  \label{Fi}
\end{equation}
Note that, the minimization problem P2 has \textcolor{black}{both} local constraints \textcolor{black}{and} global constraints. To solve it in a distributed way, we introduce a dual problem based ADMM method inspired by \cite{AProximal,Automatica}. Let $\lambda$ be the dual variable associated with the global constraints. The Lagrangian of minimization problem in step 2) of Algorithm \ref{algorithm1} is 
\begin{equation}
  \mathcal{L}_{d} = \sum_{i=1}^M F_i + \lambda^{\top} \left(\sum_{i=1}^{M} f_{i} (\hat{u}_{i})-\delta_t\right).
\end{equation}
and the dual problem is formulated as
\begin{equation}
  \max _{\lambda \geq 0} \min _{\hat{u}_i \in \mathcal{U}_i} \mathcal{L}_d(\hat{u}_i,\lambda).
  \label{e21}
\end{equation}
The dual problem is also equivalent to 
\begin{equation}
  \label{dual problem}
  \min _{\lambda \geq 0} \max _{\hat{u}_i \in \mathcal{U}_i} -\mathcal{L}_d(\hat{u}_i,\lambda):= \min _{\lambda \geq 0} \sum _{i=1}^M g_i(\lambda).
\end{equation}
where $g_i(\lambda)$ is defined as
\begin{equation}
  \label{gi}
  g_i(\lambda):=\max_{\hat{u}_i \in \mathcal{U}} -F_i - \lambda^{\top} \left(\sum_{i=1}^{M} f_{i} (\hat{u}_{i})-\delta_t\right).
\end{equation}
Problem (\ref{dual problem}) is not fully decomposable due to the common variable $\lambda$. Inspired by consensus  theory, the problem can be rewritten as
\begin{equation}
  \label{dual consensus}
  \begin{split}
    \min _{\lambda_i \geq 0} \sum_{i=1}^M g_i(\lambda _i)\quad \text{s.t.} \ &\lambda_i = \omega_{ij},\ \lambda_j=\omega_{ij}\\
    &  \forall i \in \mathcal{M}, \; (i,j) \in \mathcal{E}.
  \end{split}
\end{equation}
The augmented Lagrangian of (\ref{dual consensus}) is: 
\begin{equation}
  \begin{split}
    \mathcal{L}_q= &\sum_{i=1}^M g_i(\lambda _i)+\sum_{j\in N_i} \bigg \{\alpha_{ij}^{\top} (\lambda_i-\omega_{ij})+\beta_{ij}^\top(\lambda_j-\omega_{ij}) \\
    &+ \frac{\rho _2}{2} \left\| \lambda_i-\omega_{ij}\right \|_2^2+ \frac{\rho _2}{2}\left\| \lambda_j-\omega_{ij}\right \|_2^2 \bigg \}.
  \end{split}
\end{equation}
According to the standard ADMM, we have the following iteration steps:
\vspace{-0.3cm}
\begin{align}
  & \lambda_i ^{l+1}=\arg \min_{\lambda_i \geq 0} \mathcal{L}_p(\lambda_i, \omega_{ij}^l, \alpha_{ij}^l, \beta_{ij}^l)  \\
  & \omega_{ij}^{l+1}=\arg \min \mathcal{L}_p(\lambda_i^{l+1}, \omega_{ij}, \alpha_{ij}^l, \beta_{ij}^l)\\
  & \alpha_{i j}^{l+1}=\alpha_{i j}^{l}+\rho_2 \left(\lambda_{i}^{l+1}-w_{i j}^{l+1}\right)\\
  & \beta_{i j}^{l+1}=\beta_{i j}^{l}+\rho_2 \left(\lambda_{j}^{l+1}-w_{i j}^{l+1}\right),
\end{align}
where $l$ denotes iteration step. The results in \cite{AProximal,Automatica,Chang2014multi} show that the steps above can be simplified by substituting the following equalities:
\begin{equation}
  \label{37}
  \alpha_{ij}^l=\beta_{ji}^l, \quad \alpha_{ij}^l+\beta_{ij}^l=0, \quad \omega_{ij}^l=\omega_{ji}^l=\frac{\lambda_i^l+\lambda_j^l}{2}.
\end{equation}

\begin{algorithm}
	\caption{\textbf{Dual-Consensus ADMM}}
	\label{algorithm2}
	\begin{algorithmic}
    \STATE \textbf{Initialization:} Set $l=0$, and give $\lambda_i^0$, $\alpha_{ij}^0$, $\mathbf{u}^n$, $\theta_i^{n}$
    \STATE \textbf{Repeat}
    \STATE \textbf{1) Decision and dual variables update:} For each agent, update local variables $(\hat{u}_i^{l+1},x_i^{l+1})$ and $\lambda_i^{l+1}$ in parallel according to the following two equations
    \begin{equation*}
      \begin{aligned}
        &(x_i^{l+1},\hat{u}_i^{l+1})=\arg \min_{\substack{y_i \geq 0 \\ u_i \in \mathcal{U}_i}} J_i(x_i,\hat{u}_i)+\frac{1}{4} \big\| H_i \hat{u}_i -\frac{\delta _t}{M}\\
        & \qquad \qquad \quad\;\ -p_i^l+y_i\big\|_2^2+\frac{\rho_1}{2}\big\|\hat{u}_i-E_i \mathbf{u}_i^n+\frac{1}{\rho_1} \theta_i^n\big\|_2^2\\
        &\lambda_i^{l+1}=\frac{1}{2|N_i|\rho_2}\Big(H_i \hat{u}_i^{l+1} -\frac{\delta _t}{M} -p_i^l+y_i^{l+1}\Big).\\
      \end{aligned}
    \end{equation*}
      \STATE \textbf{2) Communication:} After local variables updating, communication manipulation is carried out between neighbors. The dual variables $\alpha_{ij}^{l+1}$ are updated by 
      \begin{equation*}
        \alpha_{ij}^{l+1}=\alpha_{ij}^l+\frac{\rho_2}{2}(\lambda_i^{l+1}-\lambda_j^{l+1}).
      \end{equation*}
      \STATE \textbf{3) Set} $l=l+1$ 
      \STATE \textbf{Until} the predifined stopping criterion is satisfied and return $\hat{u}_i^{n+1}=\hat{u}_i^{l+1}$, $x_i^{n+1}=x_i^{l+1}$.
	\end{algorithmic}
\end{algorithm}

Considering (\ref{37}), distributed ADMM steps is simplified as:
\begin{equation}
  \label{e29}
  \begin{split}
    \lambda_i^{l+1}=\arg \min \; 
    &g_i(\lambda_i)+\rho_2|{N}_i| \|\lambda_i\|_2^2+2\lambda_i \sum_{j \in N_i} \alpha_{ij}^l \\
    &- \rho_2 \lambda_i \sum_{j \in N_i} (\lambda_i^l+ \lambda_j^l)
  \end{split}
\end{equation}
\begin{equation}
  \alpha_{ij}^{l+1}=\alpha_{ij}^l+ \frac{\rho_2}{2}(\lambda_i^{l+1}-\lambda_j^{l+1}),
  \label{e32}
\end{equation}
where $|N_i|$ is the cardinality of the set ${N}_i$. By substituting (\ref{gi}) to (\ref{e29}), the above problem can be rewritten as: 
\begin{equation}
  \label{Gi}
  \begin{split}
    \min_{\lambda_i \geq 0} 
    &\max_{\hat{u}_i \in \mathcal{U}_i} \Bigg \{-J_i(x_i,\hat{u}_i) -\lambda_i^\top \left(f_i{(\hat{u}_i)}-\frac{\delta_t}{M}\right)\\
    & +(p_i^l)^\top \lambda_i -(\theta_i^n)^{\top} (\hat{u}_i-E_i \mathbf{u}^n)-\frac{\rho_1}{2} \left\|\hat {u}_i-E_i \mathbf{u}^n \right\|_2^2\\
    &+\rho_1|N_i| \|\lambda_i\|_2^2 \Bigg \}\\
    & :=\min_{\lambda_i \geq 0} \max_{\hat{u}_i \in \mathcal{U}_i}G_i(\lambda_i,\hat{u}_i),
    \vspace{-0.3cm}
  \end{split}
\end{equation}
where 
\begin{equation}
  p_i^l=2\sum_{j\in N_i}\alpha_{ij}^l-\rho_2 \sum_{j\in N_i}(\lambda_i^l+\lambda_j^l).
\end{equation}
Obviously, (\ref{Gi}) is a continuous concave-convex function, which is concave for fixed $\lambda_i$ and convex for fixed $\hat{u}_i$. By applying the min-max theorem, we have the following results
\begin{equation}
  \label{e33}
  \begin{aligned}
    &\min_{\lambda_i \geq 0} \max_{\hat{u}_i \in \mathcal{U}_i}G_i(\lambda_i,\hat{u}_i)\\
    & \quad=\max_{\hat{u}_i \in \mathcal{U}_i}\min_{\lambda_i \geq 0} G_i(\lambda_i,\hat{u}_i)\\
    & \quad=\max_{\substack{\hat{u}_i \in \mathcal{U}_i\\ y_i \geq0}}  \min_{\lambda_i } \{G_i(\lambda_i,\hat{u}_i)-y_i^\top \lambda_i\}.
  \end{aligned}
\end{equation}
where $y_i$ is the Lagrange multiplier with respect to $\lambda_i$. We can rewrite (\ref{e33}) as the following form to avoid solving the complex min-max problem.
\begin{equation}
  \label{e34}
  \begin{aligned}
    & G_i(\lambda_i,\hat{u}_i)-y_i^\top \lambda_i\\
    & = -J_i(x_i,\hat{u}_i)-\frac{1}{4} \big\| H_i \hat{u}_i -\frac{\delta _t}{M}-p_i^l+y_i\big\|_2^2\\
    &\quad +\frac{1}{|N_i|\rho_2} \Big \| \lambda_i- \frac{1}{2|N_i|\rho_2}\big(\ H_i \hat{u}_i -\frac{\delta_t}{M} -p_i^l +y_i \big)\Big\|_2^2\\
    &\quad-\frac{\rho_1}{2}\big\|\hat{u}_i-E_i \mathbf{u}_i^n+\frac{1}{\rho_1} \theta_i^n\big\|_2^2.
  \end{aligned}
\end{equation}
From (\ref{e34}), the closed-form solution of inner minimization problem can be obtained conveniently.
\begin{equation}
  \label{e36}
  \lambda_i=\frac{1}{2|N_i|\rho_2}\Big[H_i \hat{u}_i -\frac{\delta_t}{M} -p_i^l +y_i\Big].
\end{equation}
By substituting (\ref{e36}) into (\ref{e34}), the decision variables $({x}_i^{l+1},\hat{u}_i^{l+1})$ and Lagrange multiplier $y_i$ could be determined by solving the following minimization problem:
\begin{equation}
  \begin{aligned}
    (\hat{u}_i^{l+1},y_i^{l+1})
    &=\arg \min_{\substack{y_i \geq 0 \\ \hat{u}_i \in \mathcal{U}_i}} J_i(x_i,\hat{u}_i)+\frac{1}{4} \big\| H_i \hat{u}_i -\frac{\delta _t}{M}\\
    &-p_i^l+y_i\big\|_2^2 + \frac{\rho_1}{2}\big\|\hat{u}_i-E_i \mathbf{u}_i^n+\frac{1}{\rho_1} \theta_i^n\big\|_2^2.
  \end{aligned}
  \label{e43}
\end{equation}
Considering (\ref{e36}), the solution of (\ref{e29}) is 
\begin{equation}
  \lambda_{i}^{l+1}=\frac{1}{2\left|N_{i}\right| \rho_{2}}\left(H_{i} \hat{u}_{i}^{l+1}-\frac{\delta_{t}}{M}-p_{i}^{l}+y_{i}^{l+1}\right).
  \end{equation}
Hence, the problem $\mathrm{P3}$ can be solved in a distributed way, the corresponding update steps are summarized in Algorithm \ref{algorithm2}. Similar to Algorithm \ref{algorithm1}, the stopping criterion can be $|r^{l+1}|=\sum_{i=1}^M \left(H_i \hat{u}_i^{l+1}+y_i^{l+1}-\frac{\delta_t}{M} \right) \leq \bar{\epsilon}_1$, $|\hat{u}_i^{l+1}-\hat{u}_i^l|\leq \bar{\epsilon}_2$ and maximum iterations.

\begin{algorithm}[tbp]
  \caption{\textbf{Distributed Double-Consensus ADMM}}
  \label{algorithm3}
  \begin{algorithmic}
    \STATE \textbf{Initialization:} Set $n=0$, and give $x_i^{0}$, $u_i^{0}$, $\hat{u}_i^{0}$, $\lambda_i^0$, $\theta_i^0$. 
    \STATE \textbf{Repeat} 
    \FOR {all $i \in \mathcal{M}$ (in parallel) }
    \STATE \textbf{1) Local variables update:} Update local variables according to Algorithm \ref{algorithm2}.
    \STATE \textbf{2) Net variables update:}  Update net variables according to (\ref{NetVarUpdate}) with neighbors' information.
    \STATE \textbf{3) Dual variables update:} Each agent updates dual variable $\mathbf{\theta}_i$ by the following equation:
    \begin{equation*}
       \theta_i^{(n+1)}=\theta_i^{n}+ \rho _1 \left(\hat{u}_i^{n+1} - {E}_i \mathbf{u} ^{n+1}\right).
    \end{equation*}
    \ENDFOR
    \STATE \textbf{Until} the predifined stopping criterion is satisfied.
  \end{algorithmic}
\end{algorithm}

\begin{figure*}[ht]
  \centering
  \subfigure[$\Lambda=0.5$]{
    \label{FreewaySmall}
    \begin{minipage}[t]{0.3\linewidth}
    \includegraphics[width=1\linewidth]{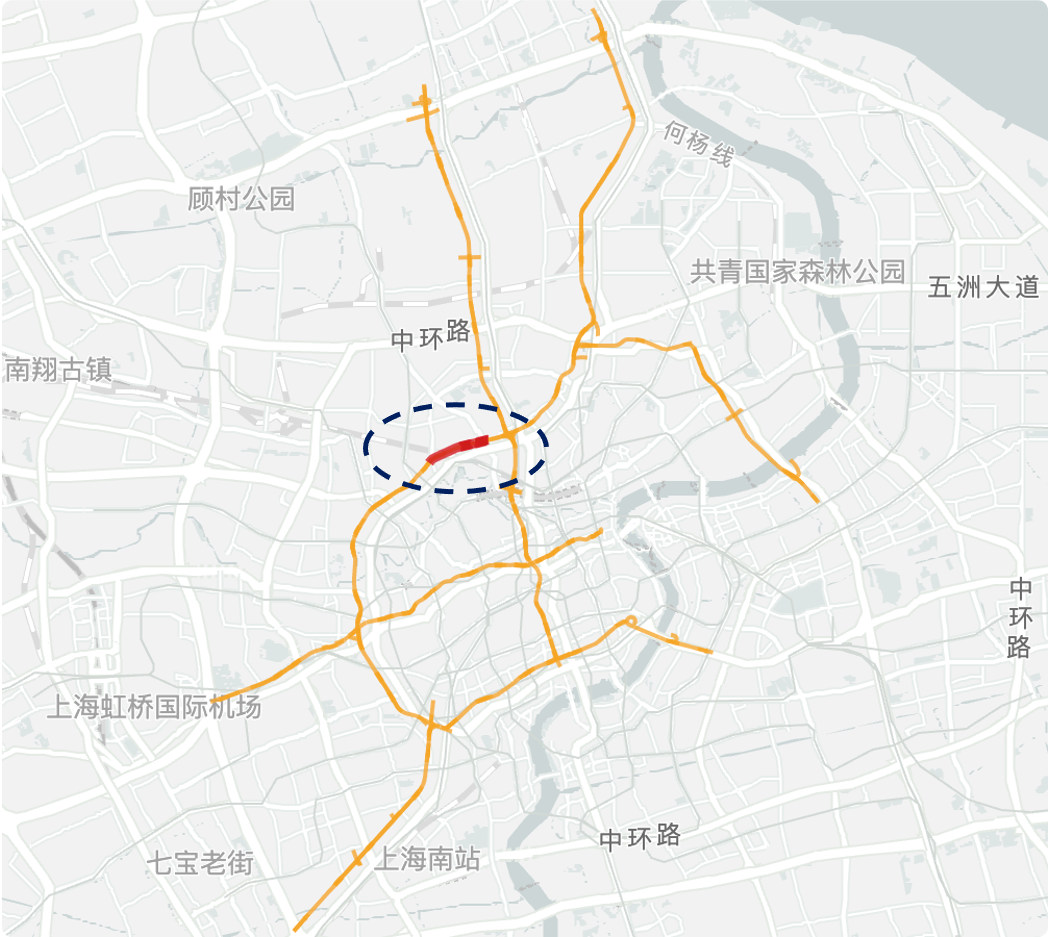}
  \end{minipage}
  }
  \subfigure[$\Lambda=0.1$]{
    \label{FreewayBig}
    \begin{minipage}[t]{0.3\linewidth}
    \includegraphics[width=1\linewidth]{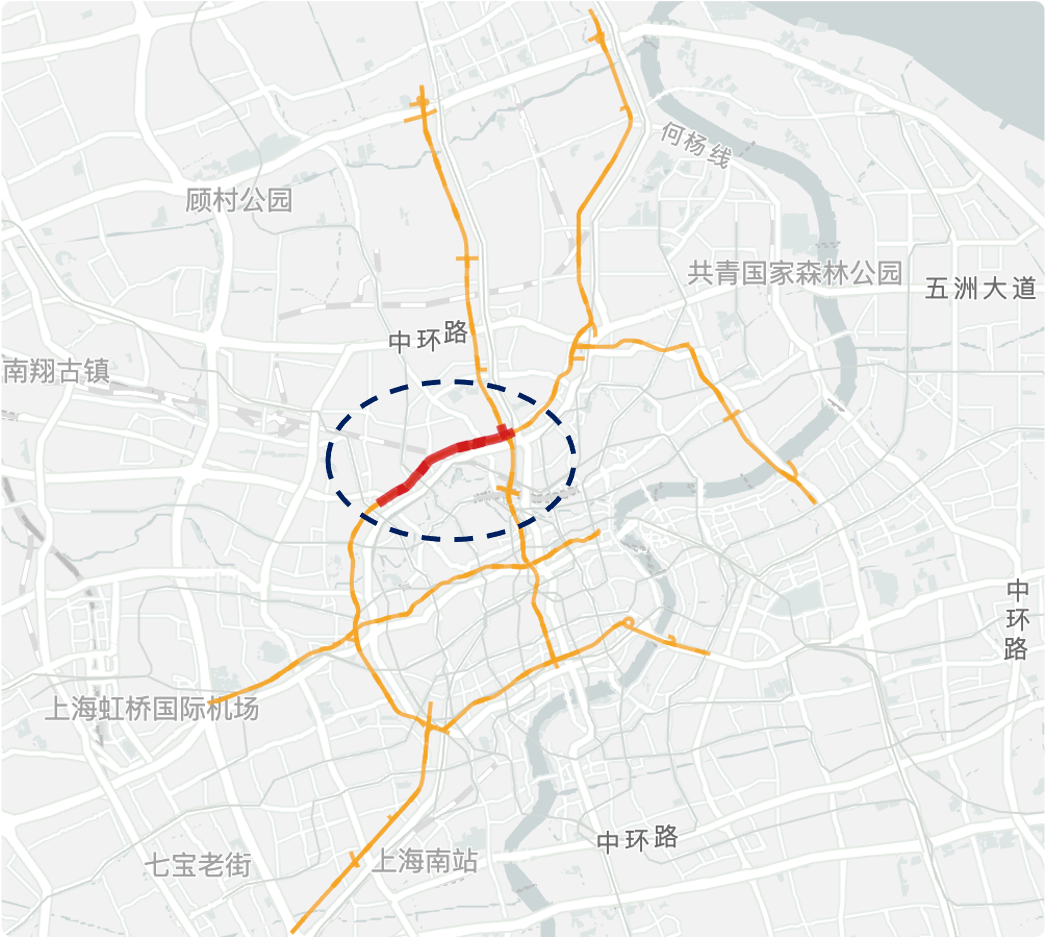}
  \end{minipage}%
  } 
  \subfigure[$\Lambda=0.05$]{
    \label{FreewayBBig}
    \begin{minipage}[t]{0.3\linewidth}
    \includegraphics[width=1\linewidth]{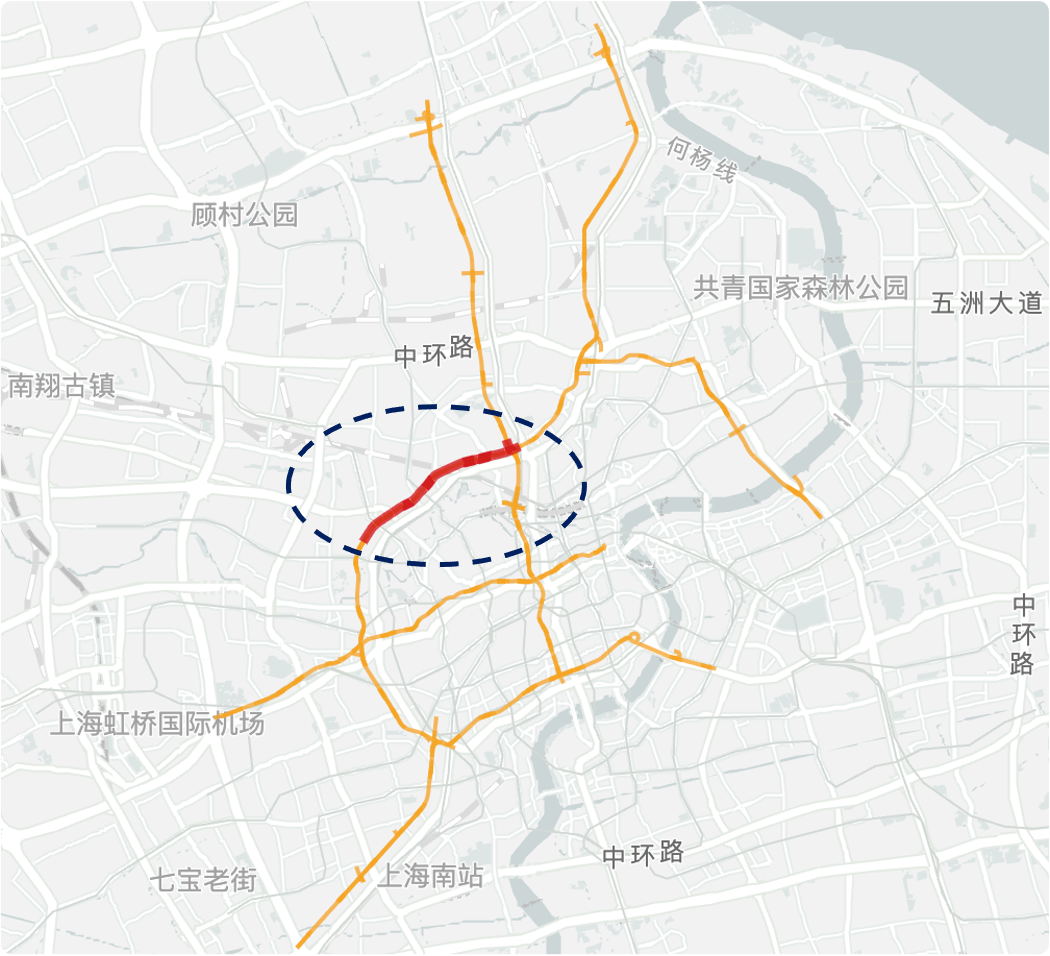}
  \end{minipage}
  }
  \vspace{-0.3cm}
    \begin{center}
      \caption{Results of spatio-temporal Lambda connectedness. The three figures are obtained under $\Lambda=0.5$, $\Lambda=0.1$, and $\Lambda=0.05$, respectively. The orange line represents normal freeway segments and red line represents the segments in \textit{potential-homogeneous-area}. 
      }
      \label{Connectedness}
    \end{center}
    \vspace{-0.6cm}
\end{figure*}

\subsection{Net Variables Update}
\label{Netupdate}
In this section, we will give the method to solve the problem P4 in step 3) of Algorithm \ref{algorithm1}. The results are given by 
\begin{equation}
  \label{NetVarUpdate}
  \begin{aligned}
    \mathbf{u}^{n+1} & = \left( \sum_{i=1}^{M} E_i^{\top} E_i\right)^{-1} \sum_{i=1}^M E_i^\top \left( \hat{u}_i^{n+1}+\frac{1}{\rho_1} \theta_i^n\right)\\
    & = \left( \sum_{i=1}^{M} E_i^{\top} E_i\right)^{-1} \sum_{i=1}^M E_i^\top \hat{u}_i^{n+1}\\
    & = \text{diag} \left( \frac{1}{|N_1|},\ldots,\frac{1}{|N_M|}\right)\sum_{i=1}^M E_i^\top \hat{u}_i^{n+1},
  \end{aligned}
\end{equation}
where the second equality is from the KKT condition of the problem P4. It is expressed as $\sum_{i=1}^{M} E_i^{\top} \theta_i^n=0$. The derivations in (\ref{NetVarUpdate}) imply that the update of net variables depends only on the information of neighbors. 

According to Algorithm \ref{algorithm1} and Algorithm \ref{algorithm2}, the original problem P1 can be solved in a fully distributed way. The corresponding results are summarized in algorithm 3, in which neighbor consensus constraint and dual consensus constraint are introduced to deal with dynamic coupling and common variable respectively. Stopping criterions are the same as aforementioned in Algorithm \ref{algorithm1}. 

\subsection{Convergence Results}
Applying Algorithm \ref{algorithm3}, the original problem P1 can be solved in a fully distributed way. The following convergence results hold
\begin{theorem}
  \label{theorem1}
  Let $\{\hat{u}_i^n,x_i^n,\theta_i^n\}$ be obtained by Algorithm \ref{algorithm3} with feasible initial conditions. Suppose $(\hat{u}_i^*,x_i^*)$ and $\theta_i^*$ are optimal solutions of the primal problem P1. If the traffic optimization objective $J_i$ is convex, the following results hold:\\
(\romannumeral1) The decision variable $\hat{u}_{i}^n$ converges to $\hat{u}_{i}^*$ for all $i \in \mathcal{M}$\\
(\romannumeral2) The residual $\left|\Delta J^{k}\right|$ converges to 0 as $k \rightarrow \infty$, where
\begin{equation}
  \begin{split}
  \Delta J^{k} &:=\sum_{i=1}^{M}\left(J_{i}\left(x_{i}, \hat{u}_{i}^k\right)-J_{i}\left(x_{i}, \hat{u}_{i}^*\right)\right).
  \end{split}
  \end{equation}
\end{theorem}
The convergence proof of \textit{Theorem} \ref{theorem1} is given in Appendix \ref{Appendix1}.

\section{Simulation}
\label{simulation}
In this section, we design several simulations based on real data to evaluate the proposed \textcolor{black}{strategies}. Firstly, we introduce the details of the urban traffic dataset \textcolor{black}{utilized in the simulations}. We then choose freeway data to verify the spatio-temporal lambda-connectedness method. Finally, the proposed distributed algorithm is applied to traffic optimization problem.

\subsection{Dataset}
In order to design convincing simulations, we use the real-world taxi data provided by Shanghai Transportation Information Center to evaluate our method. The dataset contains the traffic data of 65836 road segments during April 1-30, 2015, whose size is about 300G. Vehicle ID, GPS data, velocity, and accidents are recorded. We choose freeway data, including 720 segments, for our simulation. The summaries of the dataset are shown in Table \ref{DataSet}.

\begin{table}[htbp]
  \vspace{-0.4cm}
  \caption{Summaries of Dataset}
  \label{DataSet}
  \centering
  \begin{tabular}{c|c|c|c}
  \toprule[1pt]
  Taxi Dataset& Scale & Duration  & Contents \\  
  \cline{1-4} 
  \multirow{4}{*}{\shortstack{Taxi data,\\(Ground and \\freeway data)}} & \multirow{4}{*}{\shortstack{Freeway: 720\\Ground: 32198\\Data size: 300G}} & \multirow{4}{*}{\shortstack{2015.04.01-\\2015.04.30\;}} &\multirow{4}{*}{\shortstack{Taxi ID, GPS\\accident,\\velocity, etc.   }}   \\ 
 
  &                       &                       &                     \\ 
  \rule{0pt}{11pt}
  &                       &                       &                        \\ 
  \bottomrule[1pt]
  \end{tabular}
  \vspace{-0.6cm}
\end{table}
\subsection{Spatio-temporal Lambda Connectedness}
In this simulation, we analyze the historical data of Neihuan Road of Shanghai, China. When congestion occurs, the proposed spatio-temporal lambda connectedness method is utilized to identify the \textit{potential-homogeneous-area} \textcolor{black}{in the cloud}. The identification results are demonstrated in Fig. \ref{Connectedness}. The weighted parameters are set to $a=b=10$, and the value of $\Lambda$ is set to 0.5, 0.1, 0.05, respectively. Obviously, a smaller value of $\Lambda$ leads to a bigger range of \textit{potential-homogeneous-area}. In the scenario of $\Lambda=0.5$, the length of \textit{potential-homogeneous-area} is about 3 kilometers, and it is about 5 and 6 kilometers when $\Lambda=0.1$ and $\Lambda=0.05$. In practice, the value of $\Lambda$ can be adjusted according to the magnitude of congestion. That is, if traffic congestion is extremely severe, the administration can set $\Lambda$ a smaller value when developing traffic control strategies, which means that more segments are considered and controlled. \textcolor{black}{In this way, the congestion can be better mitigated.}

\begin{figure*}[ht]
  \centering
    \subfigure[Density under No-Control]{
    \label{Dennocontrol}
    \begin{minipage}[t]{0.3\linewidth}
    \includegraphics[width=1\linewidth]{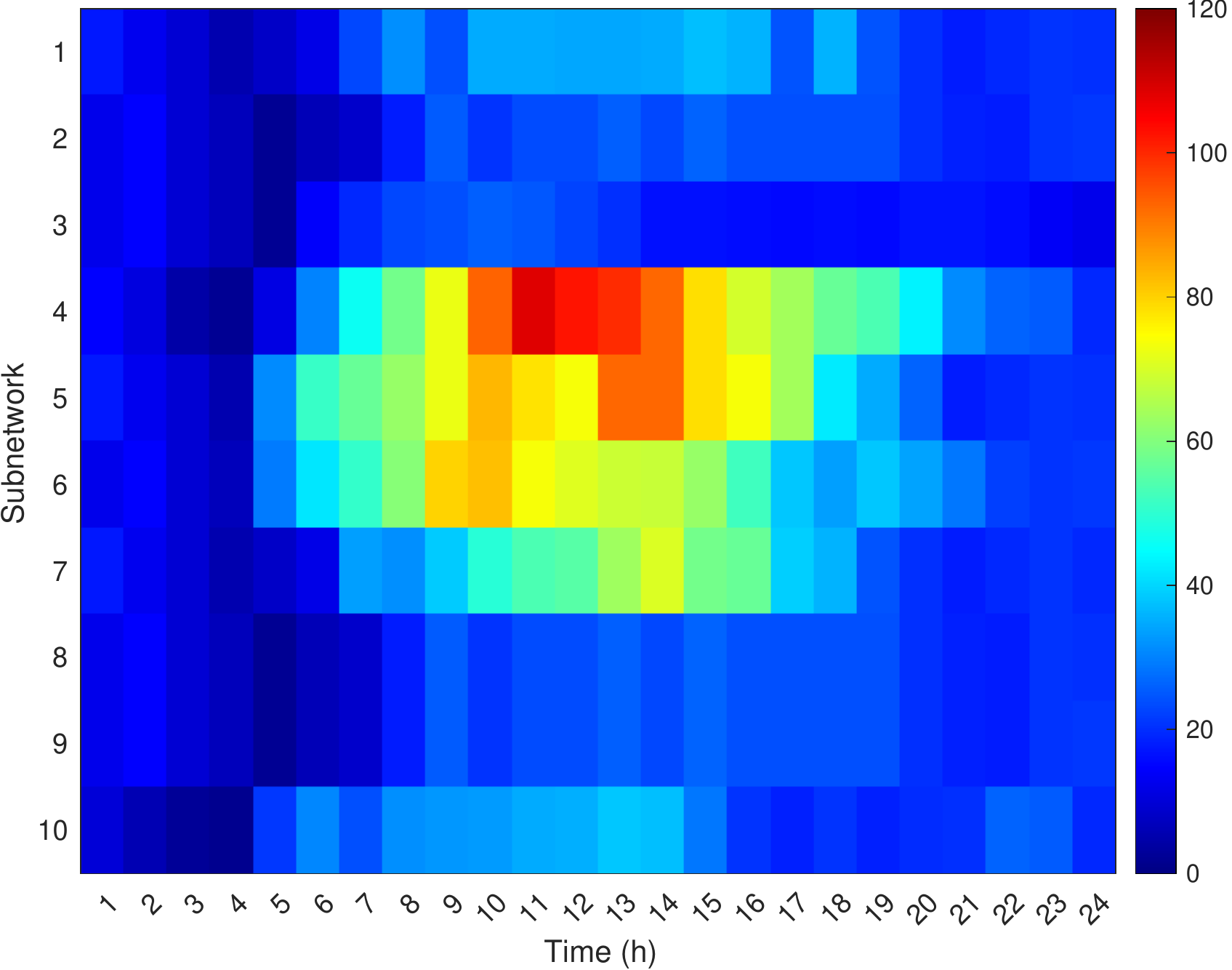}
  \end{minipage}%
  } \quad
  \subfigure[Density under ALINEA]{
    \label{DenAlinea}
    \begin{minipage}[t]{0.3\linewidth}
    \includegraphics[width=1\linewidth]{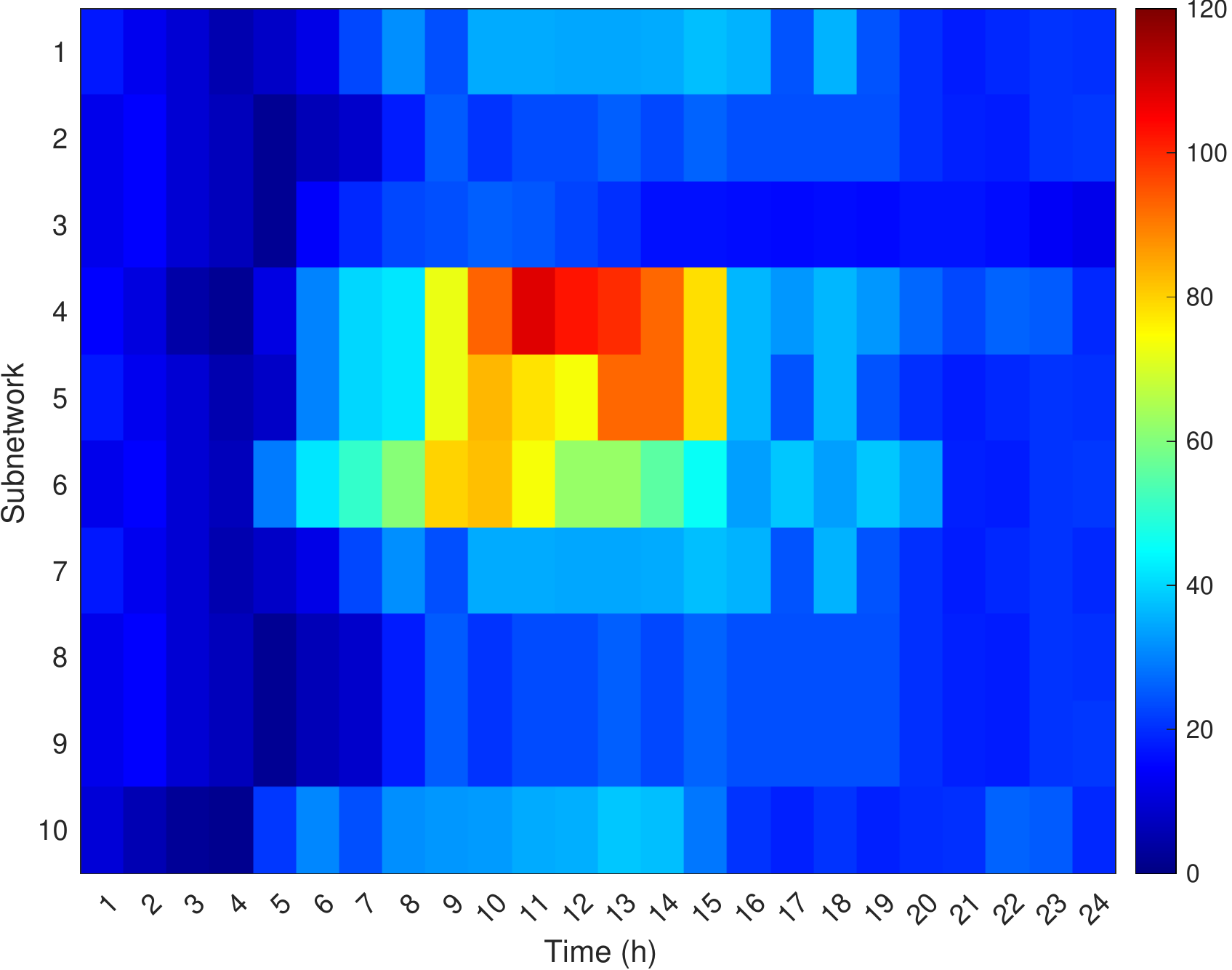}
  \end{minipage}%
  }\quad
    \subfigure[Density under $\Lambda=0.5$]{
    \label{Den3net}
    \begin{minipage}[t]{0.3\linewidth}
    \includegraphics[width=1\linewidth]{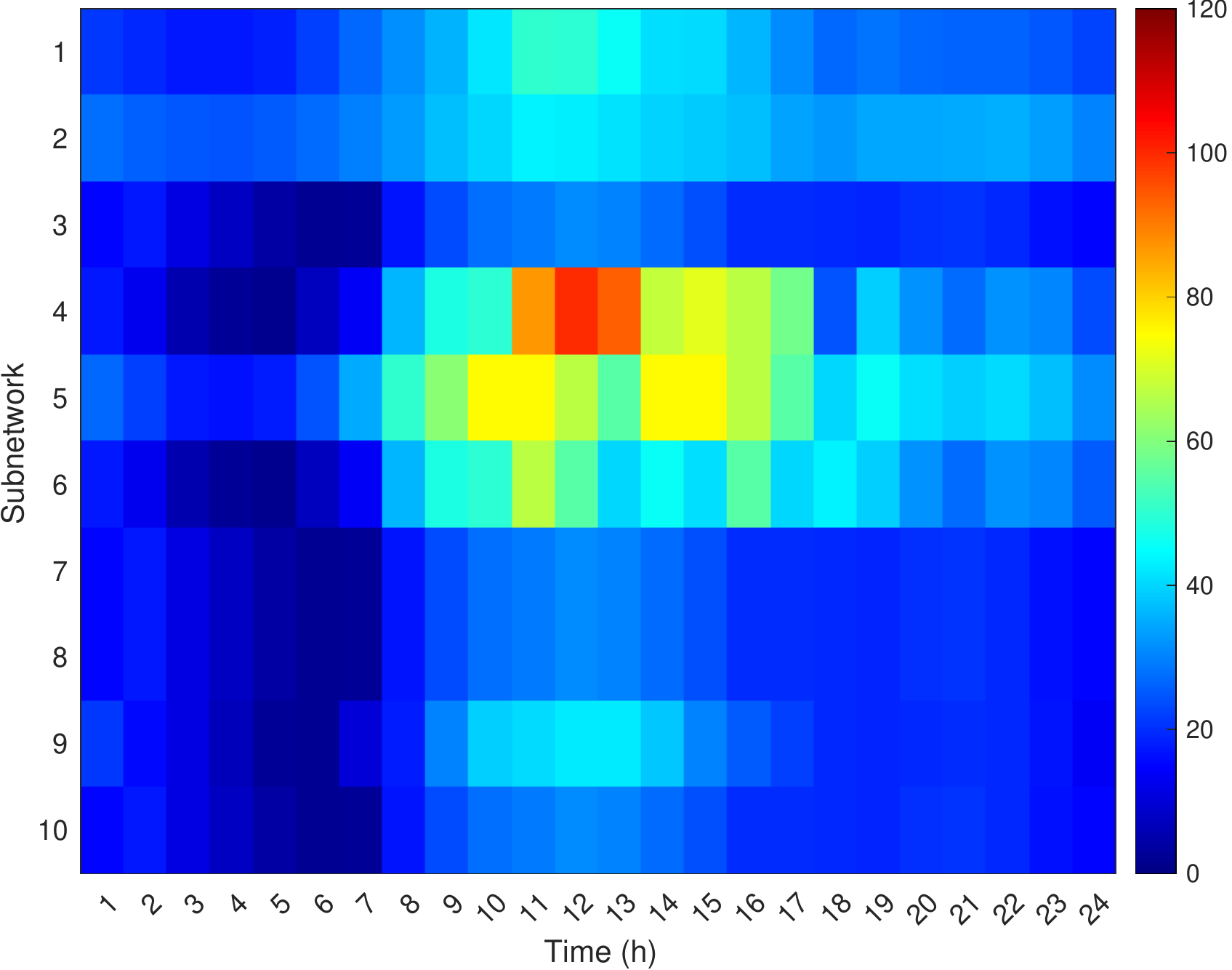}
  \end{minipage}%
  } 

  \subfigure[Density under $\Lambda=0.1$]{
    \label{Den5net}
    \begin{minipage}[t]{0.3\linewidth}
    \includegraphics[width=1\linewidth]{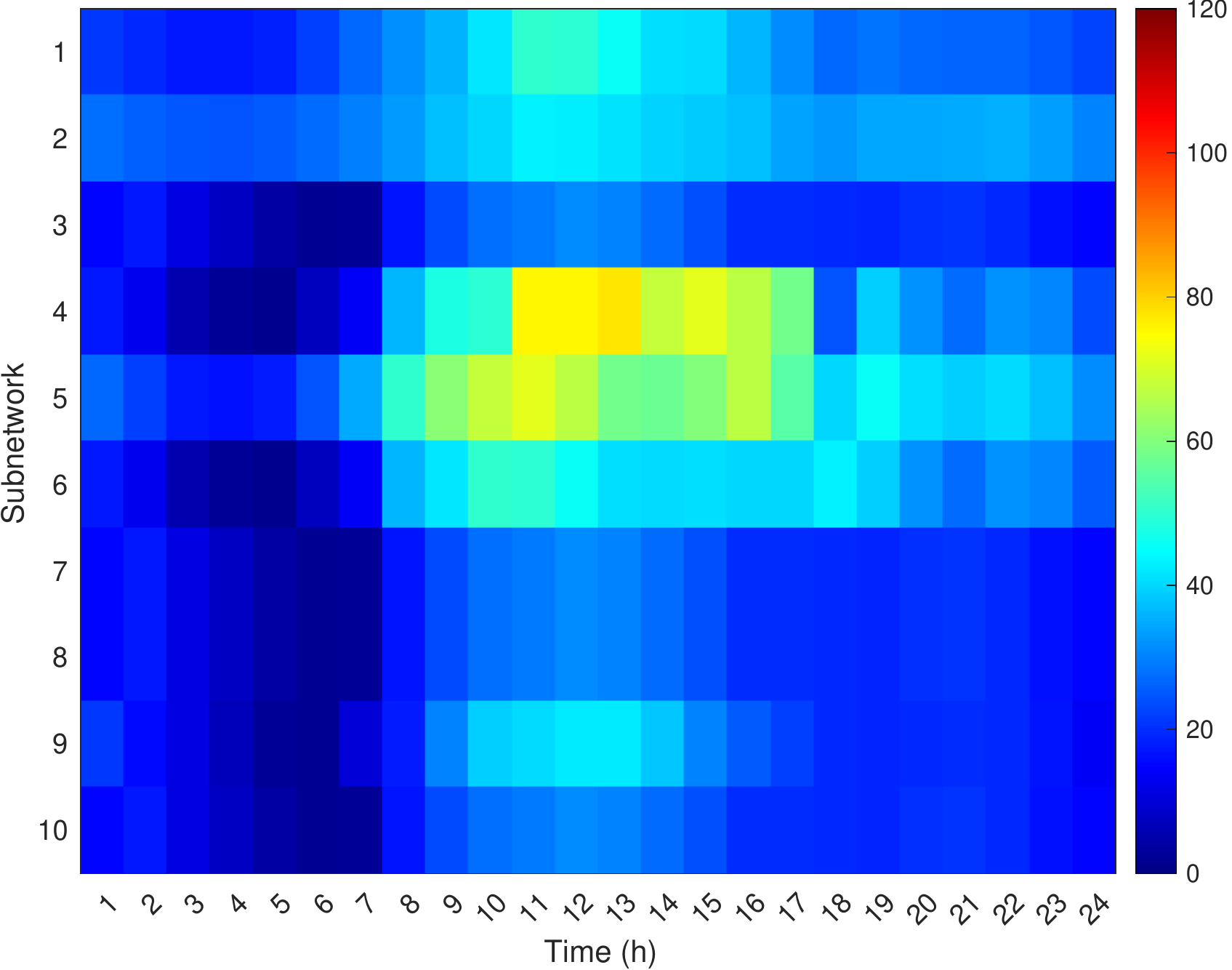}
  \end{minipage} 
  }
  \subfigure[Density under $\Lambda=0.05$]{
    \label{Den6net}
    \begin{minipage}[t]{0.3\linewidth}
    \includegraphics[width=1\linewidth]{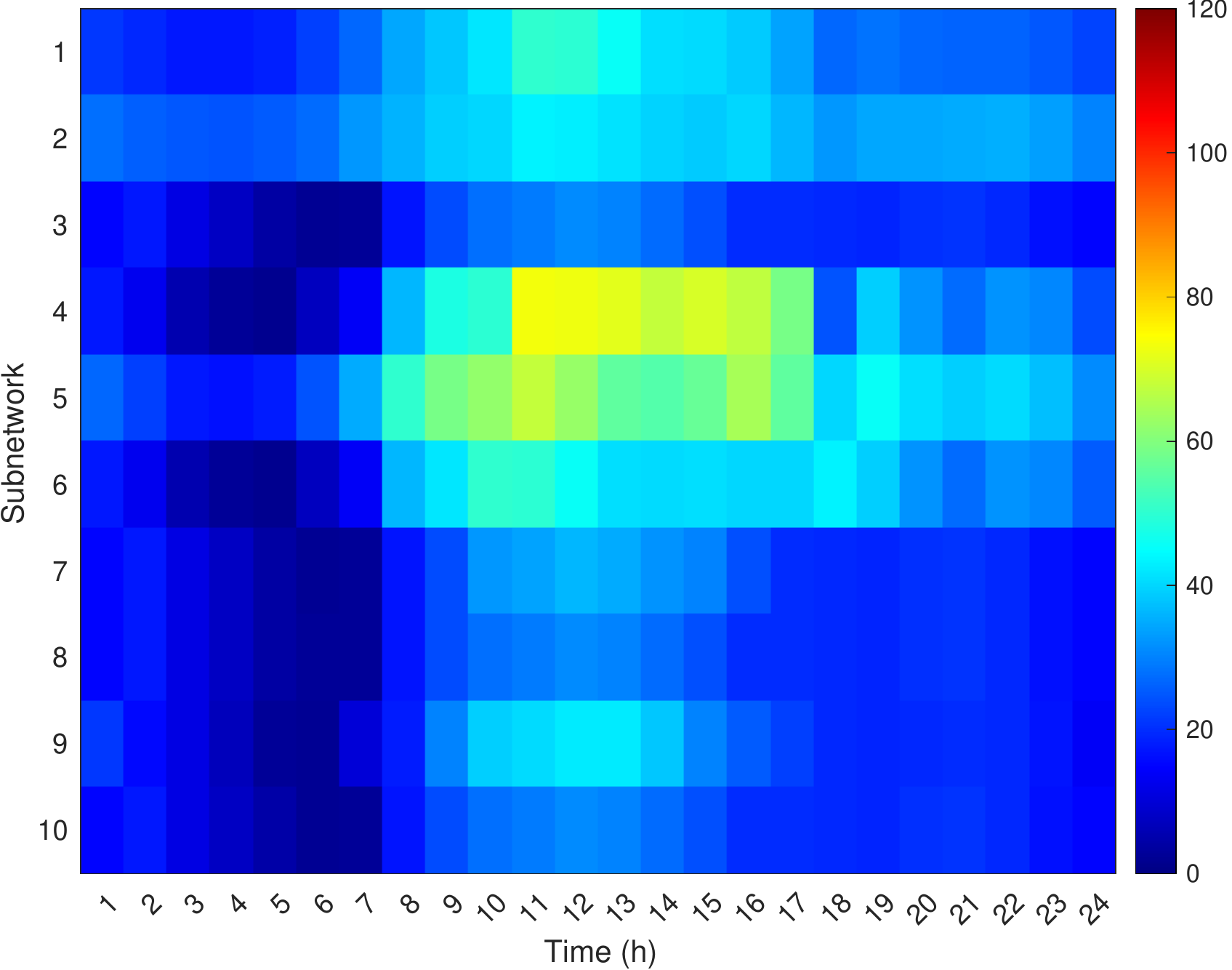}
  \end{minipage} \quad 
  }
  \caption{Traffic density fluctuation under different strategies. The five figures are the results under the strategies of no-control, ALINEA, $\Lambda=0.5$, $\Lambda=0.1$, and $\Lambda=0.05$, respectively. Fig. \ref{Dennocontrol} shows that severe congestion occurs at the segment 4, 5, and 6 during 10am-4pm. Fig. \ref{DenAlinea} demonstrates the effectiveness of ALINEA in congestion reduction. Fig. \ref{Den3net}, Fig. \ref{Den5net}, and Fig. \ref{Den6net} demonstrate that our proposed scheme can reduce congestion significantly.}
  \label{Density}
\end{figure*}

\subsection{Distributed Ramp Metering and VSL Control}
\begin{figure}[htbp]
  \centering
  \includegraphics[scale=0.5]{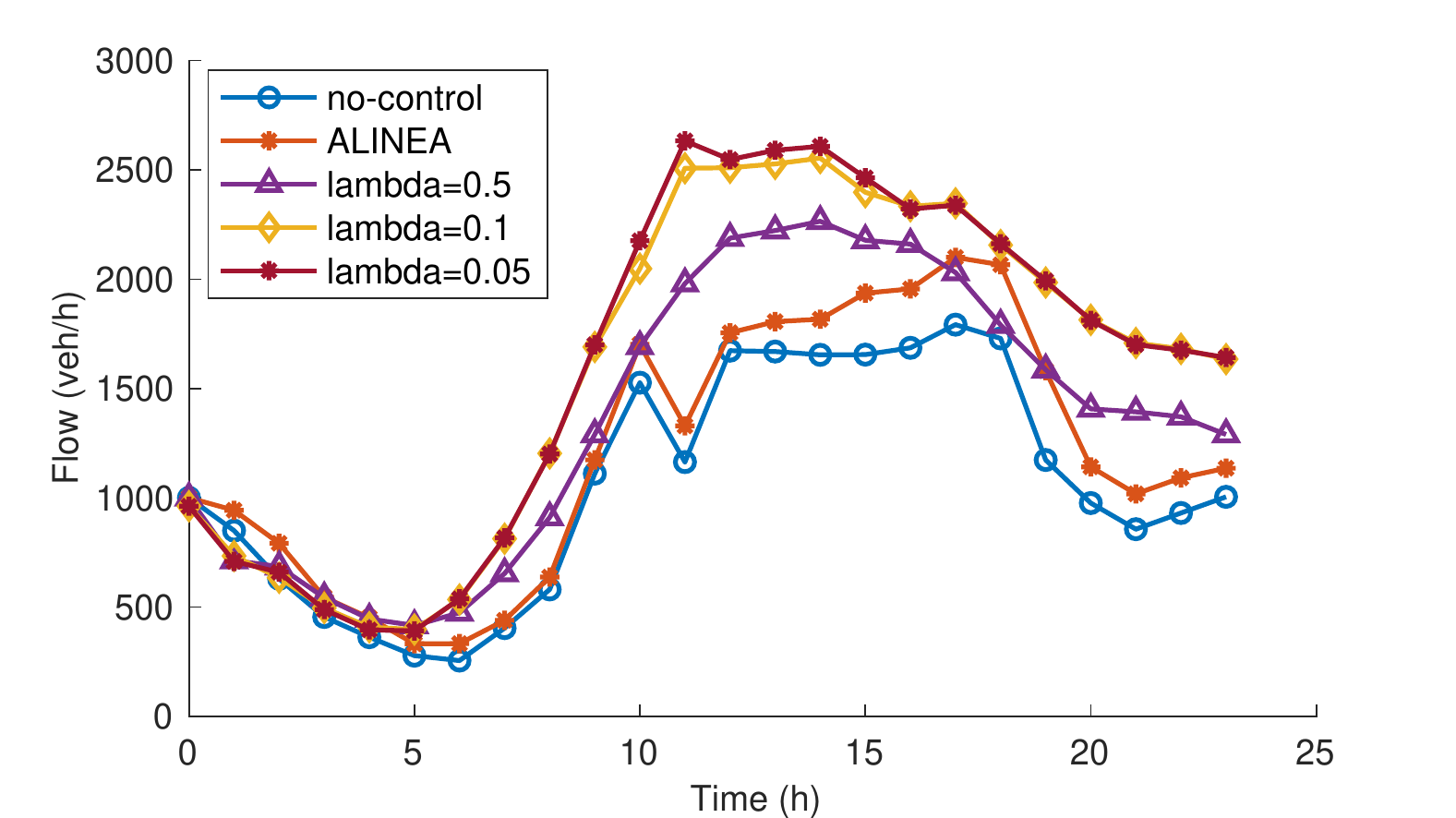}
  \vspace{-0.3cm}
  \caption{Traffic flow under different strategies.}
  \label{FlowAll}
\end{figure}

Based on the results of spatio-temporal Lambda connectedness, the segments that may be affected by congestion propagation are identified. To apply our proposed optimization algorithm, the network parameters of CTM model are first specified. Analyzing the history data, the value of CTM model parameters are given as follows: the free flow speed $v=60\ km/h$, the upper bound of traffic density $\rho_{\text{max}}=120\ veh/km$, the maximum flow $\phi_{\text{max}}=3600\ veh/h$. To investigate the performance of our proposed algorithm, we also present other control methods for comparison, \textit{e.g.}, \textit{ALINEA}. In the simulation, the whole traffic network is divided into ten subnetworks and some subnetworks are identified as \textit{potential-homogeneous-area} and controlled. The results are shown in Fig. \ref{Density} and Fig. \ref{FlowAll}. 
\vspace{-0.2cm}

Fig. \ref{Dennocontrol} shows the density fluctuation under the policy of \textit{No-Control}. In this scenario, vehicles enter the mainline freely, without ramp metering and velocity limitation. Fig. \ref{DenAlinea} demonstrates the result under classical feedback-based freeway control method, ALINEA, which is effective in mitigating congestion. Fig. \ref{Den3net}, Fig. \ref{Den5net} and Fig. \ref{Den6net} present the density fluctuation under our proposed scheme in the scenario of $\Lambda=0.5$, $\Lambda=0.1$, and $\Lambda=0.05$, respectively. In the scenario of $\Lambda=0.5$, three subnetworks are identified as \textit{potential- homogeneous-area} and controlled. It is five subnetworks in the scenario of $\Lambda=0.1$ and six subnetworks in the scenario of $\Lambda=0.05$. Comparing the five figures in Fig. \ref{Density}, it is explicit that our proposed optimization scheme is more effective in mitigating congestion. Moreover, the congestion reduction is more conspicuous in the scenario of five subnetworks controlled than three subnetworks controlled. However, smaller value of $\Lambda$ means that more subnetworks are identified as PHA and controlled. With the number of controlled subnetworks increasing, the improvement of traffic optimization performance is limited and may cause extra computation and communication costs. Hence, the trade-off between optimization performance and computation cost is essential in practice.
Results of flow fluctuation are presented in Fig. \ref{FlowAll}. The control sequence of \textit{ALINEA} is determined by real-time traffic state. Although it is effective in improving traffic throughput, this scheme has a reaction time (delay) due to the feature of feedback control. Our proposed policy minimizes the total travel time by solving a finite time horizon optimization problem, which takes future traffic demand into account. In this regard, it can avoid congestion in advance. 
\section{Conclusion}
\label[]{conclusion}
In this paper, a distributed urban freeway traffic optimization strategy considering congestion propagation has been investigated. We \textcolor{black}{firstly} propose a spatio-temporal lambda-connectedness method to \textcolor{black}{quantify the effect of congestion propagation such that the potential congested segments, \textit{i.e.,} PHA, can be identified.} \textcolor{black}{Based on identification results,}  a finite-time horizon traffic optimization problem is formulated with the dynamic capacity constraints of \textcolor{black}{PHA}. \textcolor{black}{To handle the neighbor coupling constraints and global coupling constraints within the problem, we then propose} a distributed double-consensus-based ADMM algorithm (DC-ADMM), \textcolor{black}{in which only the neighbor information is needed and the globally optimal solution is theoretically proved to be obtainable.} \textcolor{black}{The simulation results based on real data collected in Inner Ring Road, Shanghai, China, demonstrate that our proposed strategy achieves significant congestion mitigation and throughput improvement} compared with the classical strategies. \textcolor{black}{In future work, we will concentrate on traffic optimization with uncertainties in the scenario of both human-driving vehicles and automated vehicles existed.}
\appendices
\section{Proof of Theorem 1}
\label{Appendix1}
The proof is based on the convergence of Algorithm \ref{algorithm1} and Algorithm \ref{algorithm2}. We first prove that Algorithm \ref{algorithm2} converges to the optimal solution, which indicates that optimal local variables can be obtained in step 2) of Algorithm \ref{algorithm1}. In this condition, Algorithm \ref{algorithm1} converges to the optimal solution as well. As a result, \textit{Theorom} \ref{theorem1} can be proved.

The convergence of Algorithm \ref{algorithm2} is proved as follows:

We assume that the optimal solution of the problem (\ref{dual consensus}) is $\left(\hat{u}_{i}^{*}, x_{i}^{*}, y_{i}^{*}, \lambda_{i}^{*}\right)$. $\hat{u}_i^{l+1}$ and $y_i^{l+1}$ are obtained by solving (\ref{e43}). The first order optimality condition \cite{nonlinearprogramming} of (\ref{e43}) with respect to $\hat{u}_i$  and $y_i$ gives
\begin{equation}
   \begin{split}
     &\nabla_{\hat{u}_i} F_{i} \left(x_{i}, \hat{u}_{i}^{l+1}\right)^\top\left(\hat{u}_{i}^*-\hat{u}_{i}^{l+1}\right)\\
     &\qquad +(\lambda_i^{l+1})^\top H_{i}\left(\hat{u}_{i}^*-\hat{u}_{i}^{l+1}\right) \geq 0
   \end{split}
   \label{e61}
\end{equation}
\begin{equation}
  \label{ee49}
   (\lambda_i^{l+1})^{\top}\left(y_{i}^*-{y}_{i}^{l+1}\right) \geq 0
\end{equation}
where $\lambda_i^{l+1}$ is given by (\ref{e36}). Adding (\ref{e61}) and (\ref{ee49}) for $i=1,\ldots, M$ gives
\begin{equation}
  \begin{split}
    & \sum_{i=1}^M\nabla_{\hat{u}_i} F_{i} \left(x_{i}, \hat{u}_{i}^{l+1}\right)^\top \left(\hat{u}_{i}^*-\hat{u}_{i}^{l+1}\right)\\
     &\quad+\sum_{i=1}^M(\lambda_i^{l+1})^\top \Big(H_{i}\left(\hat{u}_{i}^*-\hat{u}_{i}^{l+1}\right) +y_{i}^*-{y}_{i}^{l+1} \Big)\geq 0.
   \end{split}
   \label{e63}
\end{equation}
Following the saddle point theory, the similar results can be obtained from (\ref{e21})
\begin{equation}
  \begin{split}
   &\sum_{i=1}^M\nabla_{\hat{u}_i} F_{i} \left(x_{i}, \hat{u}_{i}^{*}\right)^\top \left(\hat{u}_{i}^{l+1}-\hat{u}_{i}^{*}\right)\\
  &\quad+\sum_{i=1}^M(\lambda^{*})^\top \Big(H_{i}\left(\hat{u}_{i}^{l+1} -\hat{u}_{i}^{*}\right) +y_{i}^{l+1} -{y}_{i}^{*} \Big)\geq 0.
   \end{split}
   \label{e64}
 \end{equation}
 Adding (\ref{e63}) and (\ref{e64}) gives
 \begin{equation}
   \begin{split}
     &\sum_{i=1}^{M}\left(\nabla_{\hat{u}_{i}} F_{i}\left(x_{i}, \hat{u}_{i}^{l+1}\right)-\nabla _{\hat{u}_{i}} F_{i}\left(x_{i}, \hat{u}_{i}^{*}\right)\right)\left(\hat{u}_{i}^{l+1}-\hat{u}_{i}^{*}\right)\\
     &+\sum_{i=1}^{M}\left(\lambda_{i}^{l+1}-\lambda^{*}\right)\left(H_{i}\left(\hat{u}_i^{l+1}-\hat{u}_{i}^{*}\right)+y_{i}^{l+1}-y_{i}^{*}\right) \leq 0.
     \label{ee50}
   \end{split}
 \end{equation}
 According to the definition of $F_i(x_i,\hat{u}_i)$ in (\ref{Fi}), it is explicit that $F_i(x_i,\hat{u}_i)$ is a strong convex function with respect to $\hat{u}_i$, which implies that there exists $\mu_i \geq 0$, such that
\begin{equation}
  \label{ee53}
  \begin{split}
    &\left(\nabla_{\hat{u}_i} F_{i}\left(x_{i}, \hat{u}_{i}^{l+1}\right)-\nabla_{\hat{u}_{i}} F_{i}\left(x_{i}, \hat{u}_{i}^{*}\right)\right)^{\top} \left(\hat{u}_{i}^{l+1}-\hat{u}_{i}^{*}\right) \\
    &\qquad \qquad \qquad \qquad \qquad \qquad \quad \geq \mu_{i}\left\|\hat{u}_{i}^{l+1}-\hat{u}_{i}^{*}\right\|_{2}^{2}
  \end{split}
\end{equation}
Considering (\ref{ee53}), we can rewrite (\ref{ee50}) as
 \begin{equation}
   \begin{split}
     &  \sum_{i}^{M} \mu_{i}\left\|\hat{u}_{i}^{l+1}-\hat{u}_{i}^{*}\right\|_{2}^{2} \leq\\
     &\quad -\sum_{i=1}^{M}\left(\lambda_{i}^{l+1}-\lambda^{*}\right)^{\top}\left(H_{i}\left(\hat{u}_{i}^{l+1}-\hat{u}_{i}^{*}\right)+y_{i}^{l+1}-y_{i}^{*}\right).
   \end{split}
   \label{e66}
 \end{equation}
 From (\ref{e36}), $H_{i} \hat{u}_{i}^{l+1}+y_{i}^{l+1}$ can be rewritten as
 \begin{equation}
   H_{i} \hat{u}_{i}^{l+1}+y_{i}^{l+1}=2 \rho_{2}\left|N_i\right| \lambda_{i}^ {l+1}+\frac{\delta(t)}{M}+p_{i}^l
   \label{e67}
 \end{equation}
 Since the optimal solution of (\ref{dual consensus}) satisfies the consensus constraints $\lambda_{i}^{*}=\lambda^{*}$. From KKT condition, we have 
 \begin{equation}
   \begin{split}
     &\nabla g_{i}\left(\lambda_{i}^{*}\right)+2 \sum_{j \in N_{i}} \alpha_{ij}^*-y_{i}^{*}\\
     &\quad =-H_{i} \hat{u}_{i}^{*}+\frac{\delta_t}{M}+2 \sum_{j \in N_{i}}\alpha_{i j}^{*}-y_{i}^{*}=0.
   \end{split}
   \label{e68}
 \end{equation}
 Hence, the right side of (\ref{e66}) can be rewritten as
 \begin{equation}
  \label{ee57}
   \begin{split}
     &H_{i}\left(\hat{u}_{i}^{l+1}-\hat{u}_{i}^{*}\right)+y_{i}^{l+1}-y_{i}^{*}\\
     &\;=2 \rho_{2}\left|N_{i}\right| \lambda_{i}^{l+1}+p_{i}^{l}-2 \sum_{j \in N_{i}} \alpha_{i j}^{*}\\
     &\;=2\rho_2 |N_i|\lambda_i^{l+1}+ \sum_{j\in N_i}\left(2\alpha_{ij}^l-2\alpha_{ij}^*-\rho_2\left(\lambda_i^l+\lambda_j^l\right)\right)\\
     &\;=2 \sum_{j \in N_{i}}\left(\alpha_{i j}^{l+1}-\alpha_{i j}^{*}\right)+\rho_2 \sum_{j \in N_{i}}\left(\lambda_{i}^{l+1}+\lambda_{j}^{l+1}-\lambda_{i}^{l}-\lambda_{j}^{l}\right)
   \end{split}
 \end{equation}
 where the first equality is from (\ref{e67}) and (\ref{e68}), and the second equality is from (\ref{e32}). Substituting (\ref{ee57}) into (\ref{e66}), we have the following result 
 \begin{equation}
   \begin{split}
     &\sum_{i=1}^{M} \mu_{i}\left\|\hat{u}_{i}^{l+1}-\hat{u}_{i}^{*}\right\|_{2}^{2}\\
     &\; \leq -2 \sum_{i=1}^{M}\sum_{j\in N_i}\left(\lambda_{i}^{l+1}-\lambda^{*}\right)^{\top}\left(\alpha_{i j}^{l+1}-\alpha_{i j}^{*}\right)\\
     &\; \quad -\rho_2 \sum_{i=1}^{M} \sum_{j \in N_{i}}\left(\lambda_{i}^{l+1}-\lambda^{*}\right)^{\top}\left(\lambda_{i}^{l+1}+\lambda_{j}^{l+1}-\lambda_{i}^{l}-\lambda_{j}^{l}\right).
   \end{split}
   \label{e70}
 \end{equation}
 We consider the first term on the right side of (\ref{e70})
   \begin{align}
     &2 \sum_{i=1}^{n} \sum_{j \in N_{i}}\left(\lambda_{i}^{l+1}-\lambda^{*}\right)^{\top}\left(\alpha_{ij}^{l+1}-\alpha_{ij}^{*}\right) \notag \\
     &\quad=\sum_{i=1}^{M} \sum_{j \in N_{i}}\left(\lambda_{i}^{l+1}-\lambda^{*}\right)^{\top}\left(\alpha_{i j}^{l+1}-\alpha_{i j}^{*}\right) \notag \\
     &\qquad+\sum_{j=1}^{M} \sum_{i \in N_j}\left(\lambda_{j}^{l+1}-\lambda^{*}\right)^{\top}\left(\alpha_{j i}^{l+1}-\alpha_{j i}^{*}\right) \notag \\
     &\quad=\sum_{i=1}^{M} \sum_{j \in N_i}\left(\lambda_{i}^{l+1}-\lambda^{*}\right)^{\top}\left(\alpha_{i j}^{l+1}-\alpha_{i j}^{*}\right) \notag\\
     &\qquad-\sum_{i=1}^{M} \sum_{j \in N_i}\left(\lambda_{j}^{l+1}-\lambda^{*}\right)^{\top}\left(\alpha_{i j}^{l+1}-\alpha_{ij}^{*}\right)\notag \\
     &\quad=\sum_{i=1}^{M} \sum_{j \in N_{i}}\left(\lambda_{i}^{l+1}-\lambda_{j}^{l+1}\right)^{\top}\left(\alpha_{i j}^{l+1}-\alpha_{i j}^{*}\right) \notag \\
     &\quad=\frac{2}{\rho_2} \sum_{i=1}^{M} \sum_{j \in N_{i}}\left(\alpha_{ij}^{l+1}-\alpha_{i j }^{l}\right)^{\top}\left(\alpha_{ij}^{l+1}-\alpha_{i j}^{*}\right) \notag \\
     &=\frac{2}{\rho_2}\left(\bm{\alpha}^{l+1}-\bm{\alpha}^{l}\right)^{\top}\left(\bm{\alpha}^{l+1}-\bm{\alpha}^{*}\right) \notag \\
     &=\frac{1}{\rho_2}\left(\left\|\bm{\alpha}^{l+1}-\bm{\alpha}^{*}\right\|_{2}^{2}-\left\|\bm{\alpha}^{l}-\bm{\alpha}^{*}\right\|_{2}^{2}+\left\|\bm{\alpha}^{l+1}-\bm{\alpha}^{l}\right\|_{2}^{2}\right)
     \label{e71}
   \end{align}

   where $\bm{\alpha}^{k}$ and $\bm{\alpha}^*$ are vectors concatenated by $\alpha_{ij}^l$ and $\alpha^*$ respectively for all $i\in \mathcal{M}$ and $j \in N_i$. The last equality is obtained by the fact of
   \begin{equation}
     \begin{split}
      &\left(\mathbf{a}_{1}-\mathbf{a}_{2}\right)^{T} \mathbf{A}\left(\mathbf{a}_{1}-\mathbf{a}_{3}\right)=\frac{1}{2}\left\|\mathbf{a}_{1}-\mathbf{a}_{3}\right\|_{\mathbf{A}}^{2}\\
      &\quad+\frac{1}{2}\left\|\mathbf{a}_{1}-\mathbf{a}_{2}\right\|_{\mathbf{A}}^{2}-\frac{1}{2}\left\|\mathbf{a}_{2}-\mathbf{a}_{3}\right\|_{\mathbf{A}}^{2}.
     \end{split}
   \end{equation}
  Similarly, we have the following results for the second term on the right side of (\ref{e70}) 
 \begin{equation}
   \begin{split}
     &\rho_{2} \sum_{i=1}^{M} \sum_{j \in N_{i}}\left(\lambda_{i}^{l+1}-\lambda^{*}\right)^{T}\left(\lambda_{i}^{l+1}+\lambda_{j}^{l+1}-\lambda_{i}^{l}-\lambda_{j}^{l}\right)\\
     &=\rho_{2}\left(\boldsymbol{\lambda}^{l+1}-\boldsymbol{\lambda}^{*}\right)^{\top} \Gamma\left(\boldsymbol{\lambda}^{l+1}-\boldsymbol{\lambda}^{l}\right)\\
     &=\frac{\rho_{2}}{2}\left(\left\|\boldsymbol{\lambda}^{l+1}-\boldsymbol{\lambda}^{*}\right\|_{\Gamma}^{2}-\left\|\boldsymbol{\lambda}^{l}-\boldsymbol{\lambda}^{*}\right\|_{\Gamma}^{2}+\left\|\boldsymbol{\lambda}^{l+1}-\boldsymbol{\lambda}^{l}\right\|_{\Gamma}^{2}\right).
   \end{split}
   \label{e72}
 \end{equation}
 Substituting (\ref{e71}) and (\ref{e72}) into (\ref{e70}) yields
 \begin{equation}
   \begin{split}
     &\frac{1}{\rho_{2}}\left\|\boldsymbol{\alpha}^{l+1}-\boldsymbol{\alpha}^{*}\right\|_{2}^{2}+\frac{\rho_2}{2}\left\|\boldsymbol{\lambda}^{l+1}-\boldsymbol{\lambda}^{*}\right\|_{\Gamma}^{2}  \\
     &\; \leq \frac{1}{\rho_2}\left\|\boldsymbol{\alpha}^{l}-\boldsymbol{\alpha}^{*}\right\|_2^{2}+\frac{\rho_2}{2}\left\|\boldsymbol{\lambda}^{l}-\boldsymbol{\lambda}^{*}\right\|_{\Gamma}^{2}-\frac{1}{\rho_2}\left\|\boldsymbol{\alpha}^{l+1}-\boldsymbol{\alpha}^{l}\right\|_2^{2}\\
     &\; \quad -\frac{\rho_2}{2}\left\|\boldsymbol{\lambda}^{l+1}-\boldsymbol{\lambda}^{l}\right\|_{\Gamma}^{2} \quad-\sum_{i=1}^{M} \mu_{i} \| \hat{u}_i^{l+1}-\hat{u}_i^*\|_2^2.
   \end{split}
   \label{e73}
 \end{equation}
We define a Lyapunov function $\mathcal{P}$ as
\begin{equation}
  \mathcal{P}^l= \frac{1}{\rho_2}\left\|\boldsymbol{\alpha}^{l}-\boldsymbol{\alpha}^{*}\right\|^{2}+\frac{\rho_2}{2}\left\|\boldsymbol{\lambda}^{l}-\boldsymbol{\lambda}^{*}\right\|_{\Gamma}^{2}
\end{equation}
The inequality (\ref{e73}) can be rewritten as
\begin{equation}
  \begin{split}
    \mathcal{P}^{l+1}&=\mathcal{P}^l-\frac{1}{\rho_2}\left\|\boldsymbol{\alpha}^{l+1}-\boldsymbol{\alpha}^{l}\right\|^{2}-\frac{\rho_2}{2}\left\|\boldsymbol{\lambda}^{l+1}-\boldsymbol{\lambda}^{l}\right\|_{\Gamma}^{2} \\
    &\quad -\sum_{i=1}^{M} \mu_{i} \| \hat{u}_i^{l+1}-\hat{u}_i^*\|_2^2.
    \label{ee61}
  \end{split}
\end{equation}
Adding (\ref{ee61}) for $k=0,\dots,\infty$, we have 
\begin{equation}
  \begin{split}
    \mathcal{P}^{l+1} &\leq \mathcal{P}^0 -\sum_{k=0}^{\infty} \bigg(\frac{1}{\rho_2}\left\|\boldsymbol{\alpha}^{l+1}-\boldsymbol{\alpha}^{l}\right\|^{2}-\frac{\rho_2}{2}\left\|\boldsymbol{\lambda}^{l+1}-\boldsymbol{\lambda}^{l}\right\|_{\Gamma}^{2}\\
    &\quad -\sum_{i=1}^{M} \mu_{i} \| \hat{u}_i^{l+1}-\hat{u}_i^*\|_2^2\bigg).
  \end{split}
  \label{ee62}
\end{equation}
Note that, the left side of (\ref{ee62}) is lower bound by 0, which implies that $\boldsymbol{\alpha}^{l+1}-\boldsymbol{\alpha}^{l} \rightarrow 0$, $ \hat{u}_{i}^{l+1}-\hat{u}_{i}^{*} \rightarrow 0$ and $\boldsymbol{\lambda}^{l+1}-\boldsymbol{\lambda}^{l} \rightarrow 0$ as $l\rightarrow \infty$.


Due to the convexity of $F_i(x_i,\hat{u}_i)$, there exists $F_{i}\left(x_{i}, \hat{u}_{i}\right)-F_{i}\left(x_{i}, \hat{u}_{i}^{l+1}\right) \geq\left(\nabla_{\hat{u}_i}F_i\left(x_{i},\hat{u}_i^{l+1}\right)\right)^{\top}\left(\hat{u}-\hat{u}_i^{l+1}\right)$. Hence, (\ref{e61}) can be reformulated as
 \begin{equation}
   \begin{split}
     &F_i(x_i,\hat{u}_i^*)-F_i(x_i,u_i^{l+1})\\
     &\quad +(\lambda_i^{l+1})^\top \left(H_{i}\left(\hat{u}_{i}^{*}-\hat{u}_{i}^{l+1}\right)+y_{i}^{*}-y_{i}^{l+1}\right) \geq 0
   \end{split}
 \end{equation}
 which can be reformulated as
 \begin{equation}
  \label{ee68}
   \begin{split}
     &F_i(x_i,\hat{u}_i^*)-F_i(x_i,u_i^{l+1})\\
     &\quad +(\lambda^{*})^\top \left(H_{i}\left(\hat{u}_{i}^{*}-\hat{u}_{i}^{l+1}\right)+y_{i}^{*}-y_{i}^{l+1}\right) \\
     &\quad +(\lambda_i^{l+1}-\lambda_i^*)^\top \left(H_{i}\left(\hat{u}_{i}^{*}-\hat{u}_{i}^{l+1}\right)+y_{i}^{*}-y_{i}^{l+1}\right)\geq 0.
   \end{split}
 \end{equation}
Following the previous derivation, we have
 \begin{equation}
  \label{ee69}
   \begin{split}
     &\Delta F_i^{l+1}+\sum_{i=1}^M(\lambda^{*})^\top \left(H_{i}\left(\hat{u}_{i}^{l+1}-\hat{u}_{i}^{*}\right)+y_{i}^{l+1}-y^{*}\right)\\
     & \quad \leq -\frac{2}{\rho}\left(\boldsymbol{\alpha}^{l+1}-\boldsymbol{\alpha}^{l}\right)^{T}\left(\boldsymbol{\alpha}^{l+1}-\boldsymbol{\alpha}^{*}\right)\\
     &\qquad -\rho\left(\boldsymbol{\lambda}^{l+1}-\boldsymbol{\lambda}^{*}\right)^{T} \Gamma\left(\boldsymbol{\lambda}^{l+1}-\boldsymbol{\lambda}^{l}\right).
   \end{split}
 \end{equation}
 where $\Delta F_i^{l+1}=F_i(x_i,\hat{u}_i^{l+1})-F_i(x_i,u_i^{*})$. Combine (\ref{e68}) and the fact that $\sum_{i=1}^{M} \sum_{j \in {N}_{i}} \alpha_{i j}^{*}=0$, the second term on the left side of (\ref{ee69}) can be rewritten as 
 \begin{equation}
   \begin{split}
     &\sum_{i=1}^M(\lambda^{*})^\top \left(H_{i}\left(\hat{u}_{i}^{l+1}-\hat{u}_{i}^{*}\right)+y_{i}^{l+1}-y_{*}^{l+1}\right)\\
     &\; = \sum_{i=1}^{M}\left(\lambda^{*}\right)^{T}\Big(H_i \hat{u}_i^{l+1} +y_{i}^{l+1}-\frac{\delta_t}{M}-2 \sum_{j \in N_{i}} \alpha_{i j}^{*}\Big)\\
     &\; = \left(\lambda^*\right)^\top r^{l+1}.
   \end{split}
 \end{equation}
 Recall the problem (\ref{e21}) and the saddle point theory, we can obtain
 \begin{equation}
  \label{ee71}
   \begin{split}
   &\Delta F^{l+1}+\left(\lambda^{*}\right)^{T} r^{l+1} \\
   &\quad \geq \Delta F^{l+1}+\left(\lambda^{*}\right)^{T}\left(\sum_{i=1}^{M} H_i\hat{u}_i^{l+1}-\delta_t\right) \geq 0.
   \end{split}
   \end{equation}
 Take (\ref{ee69}) and (\ref{ee71}) into account, we have
 \begin{equation}
   \begin{split}
   &-\left(\lambda^{*}\right)^{T} r^{l+1} \leq \Delta F^{l+1} \\
   &\quad\leq -\left(\lambda^{*}\right)^{T} r^{l+1} -\frac{2}{\rho}\left(\boldsymbol{\alpha}^{l+1}-\boldsymbol{\alpha}^{l}\right)^{T}\left(\boldsymbol{\alpha}^{l+1}-\boldsymbol{\alpha}^{*}\right) \\
   &\qquad -\rho\left(\boldsymbol{\lambda}^{l+1}-\boldsymbol{\lambda}^{*}\right)^{T} \Gamma \left(\boldsymbol{\lambda}^{l+1}-\boldsymbol{\lambda}^{l}\right)
   \end{split}
  \end{equation}
  Further, the following results hold
   \begin{equation}
     \begin{split}
     &\Delta F^{l+1} \geq -\left\|\lambda^{*}\right\|\left\| r^{l+1}\right\| \\
     &\Delta F^{l+1} \leq \left\|\lambda^{*}\right\|\left\|r^{l+1}\right\|+\frac{2}{\rho}\left\|\Delta \alpha^{l+1}\right\|\left\|\alpha^{l+1}-\alpha^{*}\right\| \\
     &\quad +\rho\left\|\lambda^{l+1}-\lambda^{*}\right\|\big\|\bar{\Gamma}\big\|\left\|\Delta \lambda^{l+1}\right\|.
     \end{split}
     \end{equation}
Based on the convergence of  $\boldsymbol{\lambda}$, $\boldsymbol{\alpha}$, and $r^l$, we can conclude that $\Delta F^{l} \rightarrow 0$ as $l \rightarrow \ \infty$.

The above derivations prove that Algorithm \ref{algorithm2} converges to the optimal solution, which shows that step 2) in Algorithm \ref{algorithm1} can reach the optimal value of local variables. In this condition, the convergence of Algorithm \ref{algorithm3} can be proved. We will give the proof results in the following paragraphs.

Recall the definition of partial augmented Lagrangian, we assume that the saddle point of $L_p$ is $(\hat{u}^*,\mathbf{u}^*,\theta ^*)$. We have the following inequality 
\begin{equation}
  L_{p}\left(\hat{u}^{*}, \mathbf{u}^{*}, \theta^{*}\right) \leq L_{p}\left(\hat{u}^{n+1}, \mathbf{u}^{n+1}, \theta^{*}\right).
  \label{SaddlePoint}
\end{equation}
Considering the fact $\hat{u}_{i}^{*}-E_{i} \mathbf{u}^{*}=0$, (\ref{SaddlePoint}) can be rewritten as 
\begin{equation}
  J^* - J^{n+1}\leq \sum_{i=1}^{M} (\theta_i^*)^{\top} \mathcal{R}_i^{n+1}
  \label{e41}
\end{equation}
where $J^*$ is the optimal objective valueand, and $J^{n+1}$ is the value of $k+1$ iterations. $\mathcal{R}_i^n$ is the residual of the neighboring consensus constraint.
\begin{align}
  &J^*=\sum_{i=1}^{M} J_{i}\left(x_{i}, \hat{u}_{i}^{*}\right)\\
  &J^{n+1}=\sum_{i=1}^{M} J_{i}\left(x_{i}, \hat{u}_{i}^{n+1}\right)\\
  &\mathcal{R}_{i}^{n}=\hat{u}_{i}^{n}-E_{i} \mathbf{u}^{n} \label{residual}.
\end{align}
According to Algorithm \ref{algorithm1}, $\hat{u}_i^{n+1}$ minimizes $L_p(\hat{u},\mathbf{u}^n,\theta^n)$. The optimal condition holds
\begin{equation}
  \nabla J_{i}\left(x_{i}^{n+1}, \hat{u}_{i}^{n+1}\right)+\theta_{i}^{n}+\rho_{1}\left(\hat{u}_{i}^{n+1}-{E}_{i} \mathbf{u}^{n}\right)=0.
  \label{e45}
\end{equation}
Recall the update equation $\theta_{i}^{n+1}=\theta_{i}^{n}+\rho_1\left(\hat{u}_{i}^{n+1}-E_{i} \mathbf{u}^{n+1}\right)$, substitute this equality into (\ref{e45}) and the following result can be obtained
\begin{equation}
  \nabla J_{i}\left(x_{i}, \hat{u}_{i}^{n+1}\right)+\theta_{i}^{n+1}-\rho_{1} E_{i}\left(\mathbf{u}^{n+1}-\mathbf{u}^{n}\right)=0.
  \label{e46}
\end{equation}
The similar result with respect to $\mathbf{u}$ shows that
\begin{equation}
  -E_{i}^{\top} \theta^{n+1}=0.
  \label{e47}
\end{equation}
Results in (\ref{e46}) and (\ref{e47}) imply that $\hat{u}_i^{n+1}$ and $\mathbf{u}^{n+1}$ minimize the following two functions respectively
\begin{equation}
  J_{i}\left(x_{i}, \hat{u}_{i}\right)+\left(\theta_{i}^{n+1}+\rho_{1} E_{i}\left(\mathbf{u}^{n+1}-\mathbf{u}^{n}\right)\right)^{T} \hat{u}_{i}
  \label{e48}
\end{equation}
\begin{equation}
  -\left(\theta_{i}^{n+1}\right)^{\top} E_{i} \mathbf{u}.  \label{e49}
\end{equation}
The optimal condition of (\ref{e48}) and (\ref{e49}) satisfy
\begin{equation}
  \begin{split}
    & J_{i}\left(x_{i}, \hat{u}_{i}^{n+1}\right)+\left(\theta_{i}^{n+1}+\rho_{1} E_{i}\left(\mathbf{u}^{n+1}-\mathbf{u}^{n}\right)\right)^{\top} \hat{u}_{i}^{n+1}\\
    &\quad \leq J_{i}\left(x_{i}, \hat{u}_{i}^{*}\right)+\left(\theta_{i}^{n+1}+\rho_{1}E_{i}\left(\mathbf{u}^{n+1}-\mathbf{u}^{n}\right)\right)^{\top} \hat{u}_{i}^{*}
  \end{split}
  \label{e50}
\end{equation}
\begin{equation}
  -\left(\theta_{i}^{n+1}\right)^{\top} E_{i} \mathbf{u}^{n+1} \leq-\left(\theta_{i}^{n+1}\right)^{\top} E_{i} \mathbf{u}^{*}.
  \label{e51}
\end{equation}
Add (\ref{e50}) and (\ref{e51}) for $i=1,\dots, M$, we have
\begin{equation}
  \begin{split}
    & \sum_{i=1}^M \Big\{J_{i} \left(x_{i}, \hat{u}_{i}^{n+1}\right)-J_{i}\left(x_{i}, \hat{u}_{i}^{*}\right) \Big\}\leq \sum_{i=1}^M \Big\{-\left(\theta_{i}^{n+1}\right)^{\top} \mathcal{R}_{i}^{n+1}\\
    & \quad -\rho_{1}\left(E_{i}\left(\mathbf{u}^{n+1}-\mathbf{u}^{n}\right)\right)^{\top}\left(\mathcal{R}_{i}^{n+1}+E_{i}\left(\mathbf{u}^{n+1}-\mathbf{u}^{*}\right)\right)\Big\}.
  \end{split}
  \label{e52}
\end{equation}
Combine (\ref{e41}) and (\ref{e52}), the following result holds
\begin{equation}
  \begin{split}
    &2\sum_{i=1}^{M} \Big\{ 2 \rho_{1}\left(E_{i}\left(\mathbf{u}^{n+1}-\mathbf{u}^{n}\right)\right)^{\top}\left(\mathcal{R}_{i}^{n+1}+E_{i}\left(\mathbf{u}^{n+1}-\mathbf{u}^{*}\right)\right)\\
    & \qquad +\left(-\theta_{2}^{*}+\theta_{i}^{n+1}\right)^{\top}\Big\} \leq 0
  \end{split}
  \label{e53}
\end{equation}
Considering the second term on the left side of (\ref{e53}), we have the following derivation
\begin{equation}
  \begin{split}
    &2\left(-\theta_{i}^{*}+\theta_{i}^{n+1}\right)^{\top} \mathcal{R}_{i}^{n+1}\\
    &\; =2\left(\theta_{i}^{k}-\theta_{i}^{*}\right)^{\top} \mathcal{R}_{i}^{n+1}+2 \rho_1\left\|r_{i}^{n+1}\right\|^{2}_2\\
    &\; =\frac{2}{\rho_1}\left(\theta_{i}^{n}-\theta_{i}^{*}\right)^{T}\left(\theta_{i}^{n+1}-\theta_{i}^{n}\right)\\
    &\quad \;\ +\frac{1}{\rho_{1}}\left\|\theta_{i}^{n+1}-\theta_{i}^{n}\right\|_{2}^{2}+\rho_1\left\|\mathcal{R}_{i}^{n+1}\right\|_{2}^{2}\\
    & \; =\frac{1}{\rho_1}\left(\left\|\theta_{i}^{n+1}-\theta_{i}^{*}\right\|_{2}^{2}-\left\|\theta_{i}^{n}-\theta_{i}^{*}\right\|_{2}^{2}\right)+\rho_1\left\|\mathcal{R}_{i}^{n+1}\right\|_{2}^{2}
    \label{ee60}
  \end{split}
\end{equation}
where the first two equality is due to $\theta_{i}^{n+1}=\theta_{i}^{n}+\rho_{1} \mathcal{R}_{i}^{n+1}$ and the last equality is by rewriting $\left(\theta_{i}^{n+1}-\theta_{i}^{n}\right)=\left(\theta_{i}^{n+1}-\theta_{i}^{*}\right)-\left(\theta_{i}^{n}-\theta_{i}^{*}\right)$.
The remaining terms are
\begin{equation}
  \begin{split}
    & 2 \rho_{1}\left(E_{i}\left(\mathbf{u}^{n+1}-\mathbf{u}^{n}\right)\right)^{\top}\left(\mathcal{R}_{i}^{n+1}+E_{i}\left(\mathbf{u}^{n+1}-\mathbf{u}^{*}\right)\right)\\
    & \quad+\rho_{1}\left\|\mathcal{R}_{i}^{n+1}\right\|_{2}^{2}\\
    & =\rho_{1}\left\|\mathcal{R}_{i}^{n+1}+E_{i}\left(\mathbf{u}^{n+1}-\mathbf{u}^{n}\right)\right\|_{ }^{2}+\rho_1\left\|E_{i}\left(\mathbf{u}^{n+1}-\mathbf{u}^{n}\right)\right\|_{2}^{2}\\
    &\quad  +2 \rho_1 \left(E_{i}\left(\mathbf{u}^{n+1}-\mathbf{u}^{n}\right)\right)^{\top} E_{i}\left(\mathbf{u}^{n}-\mathbf{u}^{*}\right)\\
    & =\rho_1 \left\|\mathcal{R}_{i}^{k+1}+E_{i}\left(\mathbf{u}^{n+1}-\mathbf{u}^{n}\right)\right\|^{2}_2+ \rho_1 \left\| E_{i}\left(\mathbf{u}^{n+1}-\mathbf{u}^{*}\right)\right\|_{2}^{2}\\
    & \quad  -\rho_1\left\| E_{i}\left(\mathbf{u}^{n}-\mathbf{u}^{*}\right)\right\|_{2}^{2}
  \end{split}
  \label{ee89}
\end{equation}
where the above equalities are from 
\begin{equation}
  \begin{split}
    &\mathbf{u}^{n+1}-\mathbf{u}^{*}=\left(\mathbf{u}^{n+1}-\mathbf{u}^{n}\right)+\left(\mathbf{u}^{n}-\mathbf{u}^{*}\right)\\
    &\mathbf{u}^{n+1}-\mathbf{u}^{n}=\left(\mathbf{u}^{n+1}-\mathbf{u}^{*}\right)-\left(\mathbf{u}^{n}-\mathbf{u}^{*}\right).
  \end{split}
\end{equation}
Based on the above manipulations of (\ref{ee60}) and (\ref{ee89}), the inequality (\ref{e53}) can be reformulated as
\begin{equation}
  \begin{split}
    &\sum_{i=1}^M \Big\{\frac{1}{\rho_1}\left(\left\|\theta_{i}^{n+1}-\theta_{i}^{*}\right\|_{2}^{2}-\left\|\theta_{i}^{n}-\theta_{i}^{*}\right\|^{2}\right)\\
    &\quad +\rho_1\left(\left\| E_{i}\left(\mathbf{u}^{n+1}-\mathbf{u}^{*}\right)\right\|_{2}^{2}-\left\| E_{i}\left(\mathbf{u}^{n}-\mathbf{u}^{*}\right)\right\|_{2}^{2}\right) \\
    &\quad+\rho_1\big\|\mathcal{R}_{i}^{n+1}+ E_{i}\left(\mathbf{u}^{n+1}-\mathbf{u}^{*}\right)\big\|^{2} \Big \}\leq 0.
    \label{e57}
  \end{split}
\end{equation}
Define a Lyapunov function $\mathcal{Q}$ as 
\begin{equation}
  \mathcal{Q}^{n}=\sum_{i=1}^M \bigg(\frac{1}{\rho_1}\left\|\theta_{i}^{n}-\theta_{i}^{*}\right\|_{2}^{2}+\rho_1\left\|  E_{i}\left(\mathbf{u}^{n}-\mathbf{u}^{*}\right)\right\|_{2}^{2}\bigg)
\end{equation}
The inequality (\ref{e57}) can be rewritten as follows
\begin{equation}
  \begin{split}
    &\mathcal{Q}^{n+1} \leq \mathcal{Q}^{n}-\rho_1\sum_{i=1}^M\left\|\mathcal{R}_{i}^{n+1}+ E_{i}\left(\mathbf{u}^{n+1}-\mathbf{u}^{n}\right)\right\|^{2} \\
    &  \leq \mathcal{Q}^{n}-\rho_1 \sum_{i=1}^{M}\left\|E_{i}\left(\mathbf{u}^{n+1}-\mathbf{u}^{n}\right)\right\|^{2} -\rho_1 \sum_{i=1}^M\left\|\mathcal{R}_{i}^{n+1}\right\|_{2}^{2}
  \end{split}
  \label{e59}
\end{equation}
where the second inequality is from the fact that
\begin{equation}
  \sum_{i=1}^M -2\rho_1\left(\mathcal{R}_{i}^{n+1}\right)^{\top}\left({E}_{i}\left(\mathbf{u}^{n+1}-\mathbf{u}^{n}\right)\right) \leq 0
\end{equation}
This caondition can be proved easily by adding the following two inequalities, which is similar to the manipulation in (\ref{e48}) and (\ref{e49})
\begin{equation}
  \begin{split}
    &-\left(\theta_{i}^{n+1}\right)^{\top} {E_{i}} \mathbf{u}^{n+1} \leq-\left(\theta_{i}^{n+1}\right)^{\top} {E}_{i} \mathbf{u}^{n}\\
    &-\left(\theta_{i}^{n}\right)^{\top} {E_{i}} \mathbf{u}^{n} \leq-\left(\theta_{i}^{n}\right)^{\top} {E}_{i} \mathbf{u}^{n+1}.
  \end{split}
\end{equation}
Adding (\ref{e59}) for $n=1,\ldots,\infty$ gives
\begin{equation}
  \rho_1 \sum_{n=1}^{\infty} \sum_{i=1}^M\bigg(\left\|\mathcal{R}_{i}^{n+1}\right\|_{2}^{2}+\left\|{E}_{i}\left(\mathbf{u}^{n+1}-\mathbf{u}^{n}\right)\right\|^{2}\bigg) \leq \mathcal{Q}^0.
  \label{e60}
\end{equation}
Obviously, the left side of (\ref{e60}) is bounded by $\mathcal{Q}^0$, which implies that the following two conditions must hold as $k \rightarrow \infty$: $\mathcal{R}_i^{n+1} \rightarrow 0$ and ${E}_i(\mathbf{u}^{n+1}-\mathbf{u}^n) \rightarrow 0$. As a result, the right side of (\ref{e41}) and (\ref{e52}) converges to 0. Thus, we have $\lim_{n \rightarrow\infty} J(x_i^{n},\hat{u}_i^n)=J(x_i^*,\hat{u}_i^*)$. The convergence of Algorithm \ref{algorithm3} is proved.


\bibliography{reference}
\bibliographystyle{ieeetr}

%
\newpage
\begin{IEEEbiography}[{\includegraphics[width=1in,height=1.25in,clip,keepaspectratio]{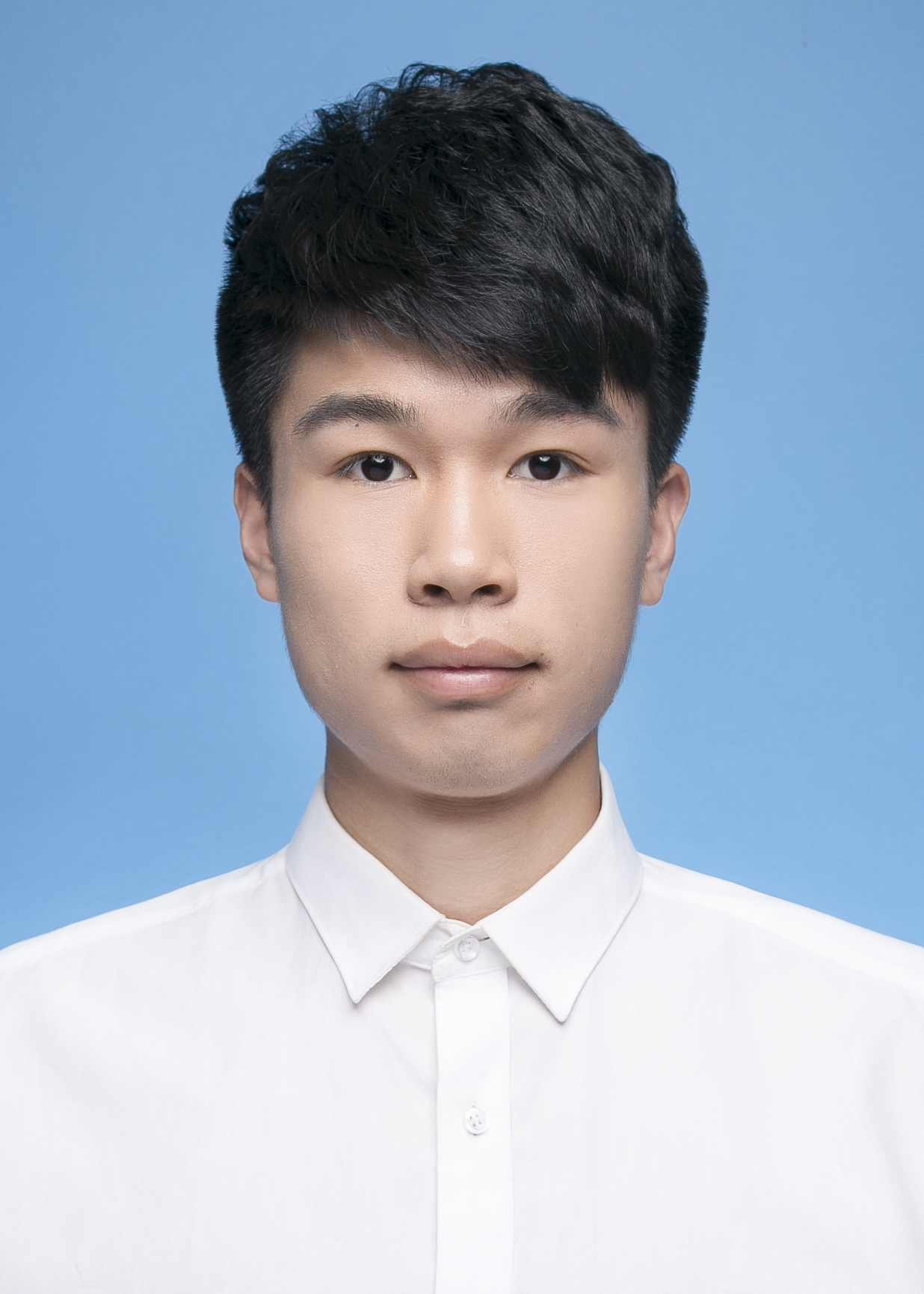}}]{Fengkun Gao}
  
 received the B.Eng. degree in the School of Control Science and Engineering, Shandong University, Jinan, China, in 2018. He is currently pursuing the Ph.D. degree at the Department of Automation, Shanghai Jiao Tong University, Shanghai, China. His current research interests include distributed traffic optimization and internet of vehicles.
\end{IEEEbiography}
\begin{IEEEbiography}[{\includegraphics[width=1in,height=1.25in,clip,keepaspectratio]{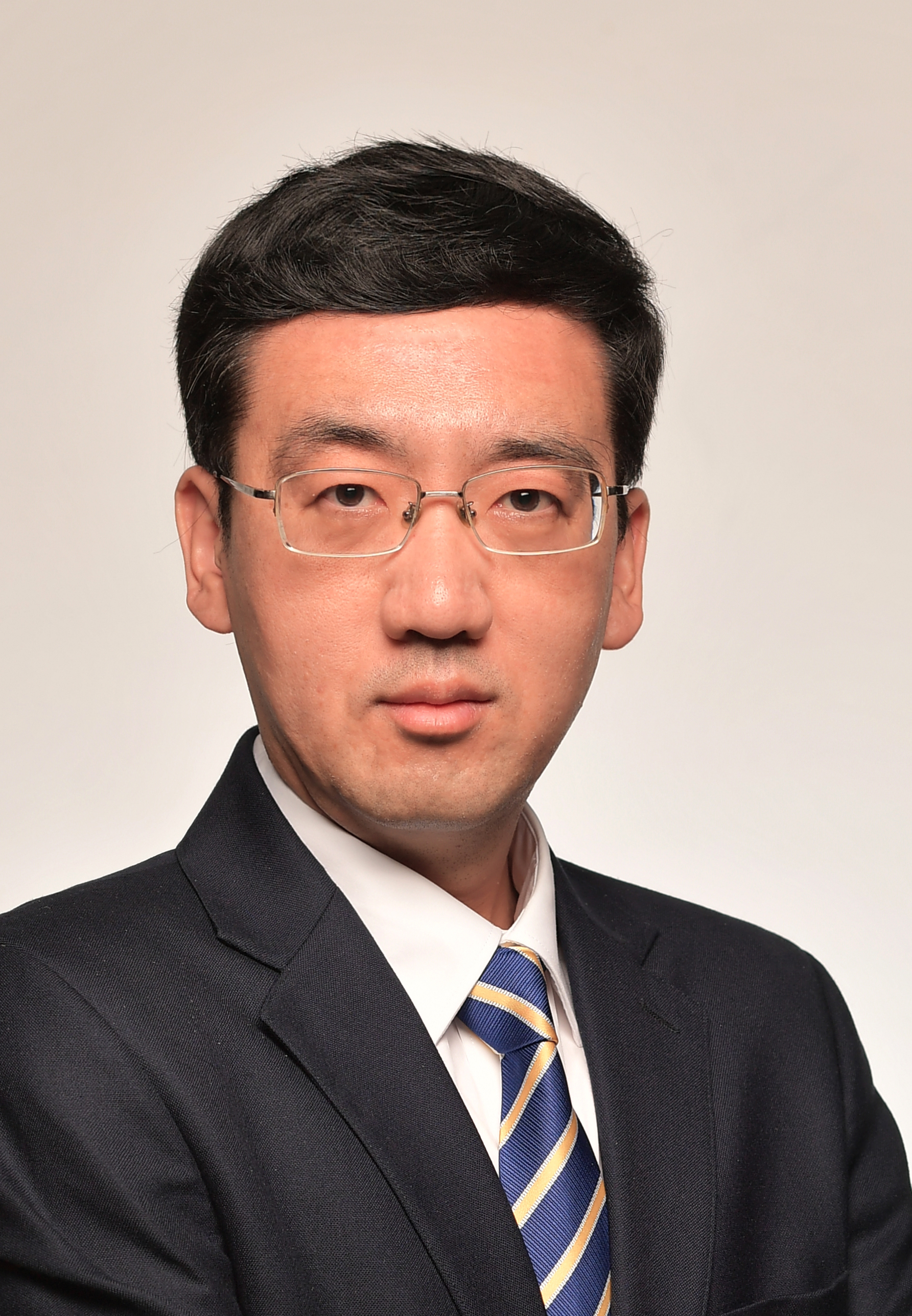}}]{Bo Yang}
(SM16) received the Ph.D. degree in electrical engineering from the City University of Hong Kong, Hong Kong, in 2009. Prior to joining Shanghai Jiao Tong University in 2010, he was a Post-Doctoral Researcher with the Royal Institute of Technology, Stockholm, Sweden, from 2009 to 2010, and a Visiting Scholar with the Polytechnic Institute of New York University in 2007. He is currently a Full Professor with Shanghai Jiao Tong University, Shanghai, China. He has authored or coauthored more than 160 articles. His research interests include control and optimization for energy networks and internet of things. Dr. Yang is on the Editorial Board of Digital Signal Processing (Elsevier) and in TPC of several international conferences. He has been the Principle Investigator in several research projects, including the NSFC Key Project. He was a recipient of the Ministry of Education Natural Science Award 2016, the Shanghai Technological Invention Award 2017, the Shanghai Rising-Star Program 2015, and the SMC-Excellent Young Faculty Award by Shanghai Jiao Tong University.

\end{IEEEbiography}

\vspace{-12 mm}
\begin{IEEEbiography}[{\includegraphics[width=1in,height=1.25in,clip,keepaspectratio]{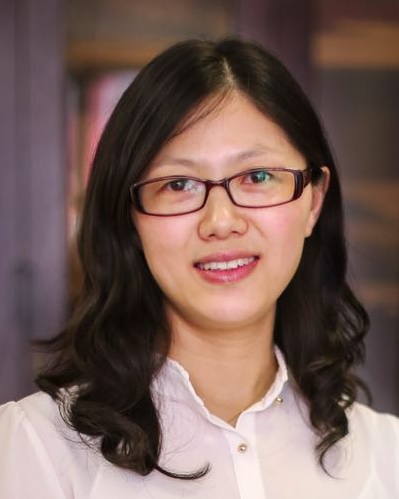}}]{Cailian Chen}
received the B. Eng. and M. Eng. degrees in Automatic Control from Yanshan University, P. R. China in 2000 and 2002, respectively, and the Ph.D. degree in Control and Systems from City University of Hong Kong, Hong Kong SAR in 2006. She has been with the Department of Automation, Shanghai Jiao Tong University since 2008. She is now a Distinguished Professor. 

Prof. Chen’s research interests include industrial wireless networks and computational intelligence, and Internet of Vehicles. She has authored 3 research monographs and over 100 referred international journal papers. She is the inventor of more than 30 patents. Dr. Chen received the prestigious "IEEE Transactions on Fuzzy Systems Outstanding Paper Award" in 2008, and 5 conference best paper awards. She won the Second Prize of National Natural Science Award from the State Council of China in 2018, First Prize of Natural Science Award from The Ministry of Education of China in 2006 and 2016, respectively, and First Prize of Technological Invention of Shanghai Municipal, China in 2017. She was honored “National Outstanding Young Researcher” by NSF of China in 2020 and “Changjiang Young Scholar” in 2015.

Prof. Chen has been actively involved in various professional services. She serves as Deputy Editor of National Science Open, and Associate Editor of IEEE Transactions on Vehicular Technology, IET Cyber-Physical Systems: Theory and Applications, and Peer-to-peer Networking and Applications (Springer). She also served as Guest Editor of IEEE Transactions on Vehicular Technology, TPC Chair of ISAS’19, Symposium TPC Co-chair of IEEE Globecom 2016, Track Co-chair of VTC2016-fall and VTC2020-fall, Workshop Co-chair of WiOpt’18.
\end{IEEEbiography}
\begin{IEEEbiography}[{\includegraphics[width=1in,height=1.25in,clip,keepaspectratio]{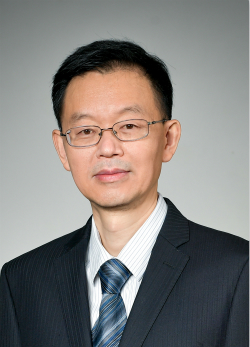}}]{Xinping Guan}
  received the B.S. degree in Mathematics from Harbin Normal University, Harbin, China, in 1986, and the Ph.D. degree in Control Science and Engineering from Harbin Institute of Technology, Harbin, China, in 1999. He is currently a Chair Professor with Shanghai Jiao Tong University, Shanghai, China, where he is the Dean of School of Electronic, Information and Electrical Engineering, and the Director of the Key Laboratory of Systems Control and Information Processing, Ministry of Education of China. Before that, he was the Professor and Dean of Electrical Engineering, Yanshan University, Qinhuangdao, China.

  Dr. Guan’s current research interests include industrial cyber-physical sys tems, wireless networking and applications in smart factory, and underwater networks. He has authored and/or coauthored 5 research monographs, more than 270 papers in IEEE Transactions and other peer-reviewed journals, and numerous conference papers. As a Principal Investigator, he has finished/been working on many national key projects. He is the leader of the prestigious Innovative Research Team of the National Natural Science Foundation of China (NSFC). Dr. Guan received the First Prize of Natural Science Award from the Ministry of Education of China in both 2006 and 2016, and the Second Prize of the National Natural Science Award of China in both 2008 and 2018. He was a recipient of IEEE Transactions on Fuzzy Systems Outstanding Paper Award in 2008. He is a National Outstanding Youth honored by NSF of China, Changjiang Scholar by the Ministry of Education of China and State-level Scholar of New Century Bai Qianwan Talent Program of China.

\end{IEEEbiography}
\begin{IEEEbiography}[{\includegraphics[width=1in,height=1.25in,clip,keepaspectratio]{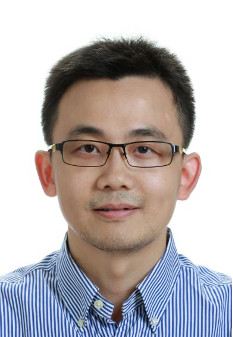}}]{Yang Zhang}
received the B.Eng. degree in Electronics and Information Technology, and the M.Eng. degree in Signal and Information Processing from Shandong University, Ji Nan, P.R. China, in 2001 and 2004, respectively, and the Ph.D. degree in Pattern Recognition and Intelligent Systems from Shanghai Jiao Tong University, Shanghai, P.R. China, in 2009. After that, he joined Shanghai Municipal Transportation Information Center as an engineer. Dr. Zhang has worked actively on applications of techniques such as artificial intelligence, and big data, etc. to Intelligent Transportation Systems (ITS) and Urban Fine Management. He received the Shanghai Rising-Star Program in 2014, and Excellent Youth Foundation of Shanghai Construction and Transportation in 2018. He is now an advanced engineer and deputy director with Shanghai Municipal Urban-Rural Construction and Transportation Development Institute. He is also a visiting professor with Institute of Smart City, Shanghai University (2018-2021).
\end{IEEEbiography}





\end{spacing}
\end{document}